\documentclass[%
 reprint,
 superscriptaddress,
 amsmath,amssymb,
 aps,
 pre,
 floatfix,
]{revtex4-2}

\usepackage{graphicx}
\usepackage{dcolumn}
\usepackage{bm}
\usepackage{hyperref}
\usepackage[dvipsnames]{xcolor}
\usepackage{float}
\hypersetup{
    colorlinks = true,
    linkcolor = blue,
    anchorcolor = blue,
    citecolor = blue,
    filecolor = blue,
    urlcolor = black
}

\begin{document}

\preprint{APS/123-QED}

\title{Phase-field model of alloy solidification far from chemical equilibrium at the solid-liquid interface}

\author{Kaihua Ji}
 \email{ji5@llnl.gov}
\affiliation{%
Physics Department and Center for Interdisciplinary Research on Complex Systems, Northeastern University, Boston, Massachusetts 02115, USA
}
\affiliation{%
Materials Science Division, Lawrence Livermore National Laboratory, Livermore, CA, 94550, USA
}%
\author{Mingwang Zhong}
 \email{m.zhong@northeastern.edu}
\author{Alain Karma}%
 \email{Corresponding author; a.karma@northeastern.edu}
\affiliation{%
Physics Department and Center for Interdisciplinary Research on Complex Systems, Northeastern University, Boston, Massachusetts 02115, USA
}%

\date{\today}

\begin{abstract}
We further develop a recently introduced phase-field model of far-from-equilibrium alloy solidification under additive manufacturing conditions [K.\ Ji et al., Phys.\ Rev.\ Lett.\ \textbf{130}, 026203 (2023)]. This model utilizes enhanced solute diffusivity within the spatially diffuse interface region to quantitatively capture solute trapping with a larger interface width, thereby making simulations on experimentally relevant length and time scales computationally feasible. The main developments presented here include testing the robustness of different variational formulations, extending the model to concentrated alloys by incorporating solid and liquid free energies from thermodynamic databases, as illustrated for hypoeutectic Al-Ag alloys with CALPHAD, extending convergence tests as a function of interface width to 3D, and carrying out simulations in both 2D and 3D to examine existing theories of microstructure development. Our results indicate that the simplest variational formulation that interpolates the bulk free-energy density between its solid and liquid forms is the most robust. Remarkably, for hypoeutectic Al-Ag alloys, this formulation yields a high-velocity nonequilibrium phase diagram that is independent of interface width, thereby demonstrating that the framework of enhanced solute diffusivity can be non-trivially extended to concentrated alloys. Other variational formulations have restricted ranges of materials or processing parameters that can be reliably modeled. We use 2D simulations to construct high-velocity microstructure selection maps for dilute Al-Cu alloys. The results validate the important role of latent heat rejection at the interface and extend the limited predictions of linear stability analysis [A.\ Karma and A.\ Sarkissian, Phys.\ Rev.\ E \textbf{47}, 513 (1993)] and sharp-interface 1D simulations to fully nonlinear regimes. Furthermore, 3D simulations, carried out using a computationally tractable axisymmetric cellular/dendritic interface shape, demonstrate a good convergence similar to that observed in 2D as a function of interface width. Full 3D simulations, in turn, reveal that the standard theory of absolute stability is a good predictor of the upper critical velocity beyond which steady-state growth becomes unstable, despite the different morphological manifestations of this instability in 2D and 3D.
\end{abstract}

\maketitle


\section{Introduction}

Alloy microstructures develop during solidification processes that range from conventional casting to advanced additive manufacturing techniques, where the solid-liquid interface velocity $V$ spans six orders of magnitude from micron/s to m/s \cite{kurz1989fundamentals,dantzig_solidification_2016}. Understanding the formation and evolution of these microstructures is essential for controlling the properties of the manufactured alloys, and it requires multiscale computational modeling. The phase-field (PF) method has emerged as a major contributor for modeling and predicting mesoscale microstructures during solidification processes \cite{boettinger_phase-field_2002,steinbach_phase-field_2009,karma_atomistic_2016,kurz_progress_2020,tourret_phase-field_2022}. This method describes a two-phase (solid-liquid) system by a scalar PF that takes constant values in each of the bulk phases and varies smoothly across interfaces of a characteristic thickness $W$. This scalar field can be coupled to the local thermodynamic state variables, such as temperature and solute concentration, and avoids the need for explicit front tracking. 
A number of PF studies have successfully predicted and elucidated microstructure patterns at length scales $\sim 100$ $\mu$m observed in experiments, predominantly under slow solidification conditions where the interface is close to thermodynamic equilibrium \cite{haxhimali_orientation_2006,dantzig_dendritic_2013,bergeon_spatiotemporal_2013,clarke_microstructure_2017,song_thermal-field_2018,wang_modeling_2020,mota_influence_2023,song_cell_2023}. Despite the physical solid-liquid interface thickness being around $W_0 \sim 1$ nm—significantly smaller than the scale of microstructural patterns—PF simulations can be conducted at experimentally relevant length and times scales by upscaling the diffuse interface thickness \cite{karma_quantitative_1998,Karma2001,echebarria_quantitative_2004,folch_quantitative_2005,plapp_unified_2011,boussinot_achieving_2014}, such that $W \gg W_0$. Techniques like the anti-trapping current help eliminate spurious effects from this upscaling \cite{Karma2001,echebarria_quantitative_2004}, ensuring local thermodynamic equilibrium at the interface. As a result, quantitative agreements between PF simulations and well-controlled experiments can be established \cite{bergeon_spatiotemporal_2013,song_thermal-field_2018,song_cell_2023}. Yet, modeling conditions far from equilibrium, as seen in rapid solidification processes like additive manufacturing, introduces unique challenges, encompassing two main aspects detailed below.

\begin{figure}[htbp!]
\includegraphics[scale=1]{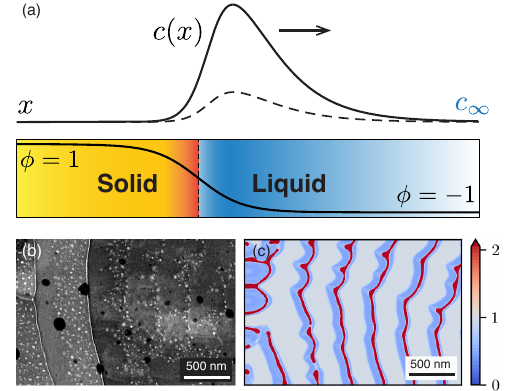}
\centering
\caption{
(a) Schematic representations of one-dimensional concentration $c(x)$ and PF $\phi(x)$ profiles during rapid solidification of binary alloys, where the positive $x$ directs to the right side of the figure. $c_{\infty}$ is the nominal concentration away from the interface. The solid and dashed $c(x)$ curves represent lower and higher solidification velocities, respectively. (b)-(c) Comparison of banded microstructures for Al-9 wt.\% Cu from the late stage of a thin-film resolidification experiment \cite{mckeown_time-resolved_2016} and a 2D PF simulation with thermal diffusion \cite{ji_microstructural_2023} ($\tilde{c}=c/c_\infty$ color map). Reproduced with permissions.
\label{fig:illustration}
}
\end{figure}

The first challenge lies in modeling an out-of-equilibrium interface with complex morphologies, where nonequilibrium effects become dominant during rapid solidification. As the solid-liquid interface propagates at an increased velocity, the solid phase will trap excess solute, resulting in a reduction of the peak concentration as illustrated in the schematics of Fig.~\ref{fig:illustration}(a). This phenomenon, known as solute trapping, occurs when the diffuse solid-liquid interface region fully crystallizes before solute partitioning is complete, where the former takes a time $\sim W/V$ for a diffuse interface moving at $V$, and the latter is established on the characteristic time $\sim W^2/D_l$ for solute atoms of diffusivity $D_l$ (in the liquid) to diffuse across the interface \cite{aziz_interface_1996}. When these two time scales become comparable, i.e., $V$ is close to a diffusive speed $V_d \equiv D_l/W$, the interface dynamics is largely controlled by the nonequilibrium effects. This is especially true for the range of $V$ approaching or exceeding the so-called absolute stability limit $V_a$, beyond which the solid-liquid interface is theoretically expected to become morphologically stable \cite{mullins_stability_1964,trivedi_morphological_1986}, when the stabilizing effect of surface tension surpasses the destabilizing effect of the solutal diffusion field. 
Except for very dilute alloy concentrations \cite{ludwig_direct_1996,boettinger_simulation_1999}, a smooth transition from cellular-dendritic to planar front growth is not typically observed. Instead, ``banded microstructures'' form [one example given in Fig.~\ref{fig:illustration}(b)], which consist of an alternation between bands exhibiting a microsegregation pattern characteristic of dendritic array growth, and microsegregation-free bands characteristic of planar front growth \cite{boettinger_effect_1984,zimmermann_characterization_1991,gremaud_banding_1991,carrard_about_1992,gill_rapidly_1993,gill_rapidly_1995,gremaud_microstructure_1990,kurz_banded_1996,mckeown_situ_2014,mckeown_time-resolved_2016}. Linear stability analyses that incorporate nonequilibrium effects have predicted the existence of oscillatory modes of the planar interface \cite{coriell_oscillatory_1983,merchant_morphological_1990,karma_interface_1993}. Nonlinear oscillations of the planar interface have indeed been shown to exist, but are insufficient to fully describe banding. Approximate analytical models of banding have also been developed, but assume instantaneous transitions between steady-state dendritic and planar front growth \cite{gremaud_banding_1991,carrard_about_1992,kurz_banded_1996}. They also neglect latent-heat rejection found to strongly affect planar front oscillations \cite{karma_dynamics_1992,karma_interface_1993}. 
While these studies have provided useful insights, a quantitative modeling approach is essential to address basic questions about dendritic array pattern stability and the underlying mechanisms of banding. To achieve this objective, the computational modeling needs to incorporate alloy phase diagram, relevant thermal conditions, energetic and kinetic anisotropies of a solid-liquid interface, and importantly, nonequilibrium solute trapping effects over a broad velocity range, extending beyond the absolute stability limit.

The second challenge revolves around carrying out simulations on experimentally relevant length and time scales. While PF models have been shown to reproduce solute trapping properties for a physical choice of interface thickness $W_0 \sim$ 1 nm in one dimension \cite{ahmad_solute_1998,karma_phase-field_2003,danilov_phase-field_2006,galenko_solute_2011,steinbach_phase-field_2012}, two-dimensional (2D) and three-dimensional (3D) simulations at the microstructural pattern scale generally necessitate the choice of an increased interface thickness in the PF model, i.e., $S \equiv W / W_0 \gg 1$. This selection, however, induces spurious excess trapping. Given that the diffusive speed is inversely related to the chosen interface thickness $S$ through the relation $V_d \sim D_l / W \sim V_d^0 / S$ (where $V_d^0 \equiv D_l / W_0$ is a constant), the excess trapping induced by $S \gg 1$ can profoundly influence interface dynamics, especially when nonequilibrium effects become dominant with $V \sim V_d$. In the context of slow solidification, this complication has been circumvented by introducing an anti-trapping current \cite{Karma2001,echebarria_quantitative_2004}, which effectively eliminates excess solute trapping, thereby restoring local equilibrium at the interface. The form of this current has been modified to also account for a moderate departure from equilibrium below the absolute stability limit \cite{pinomaa_quantitative_2019}. However, the anti-trapping current strategy is not suited to quantitatively describe a strong departure from chemical equilibrium since its magnitude increases linearly with velocity, thereby preventing complete trapping to occur in a velocity range $V\gg V_d$ where complete trapping should occur. Therefore, to describe far-from-equilibrium phenomena such as banding, a different strategy is needed to quantitatively describe solute trapping across a wide velocity range, spanning $V\sim V_d$ to $V\gg V_d$.

In our recent study \cite{ji_microstructural_2023}, a PF model was developed to address the aforementioned dual challenges. This model incorporates well-known nonequilibrium effects, including solute trapping characterized by velocity-dependent forms of the partition coefficient $k(V)$ and liquidus slope $m(V)$, as well as interface kinetics. A computationally tractable choice of $W$ is used to model the solidification of a dilute Al-Cu alloy under far-from-equilibrium conditions. To compensate for the excess solute trapping, the model adopts a novel approach by enhancing the solute diffusivity $D(\phi) \equiv D_l q(\phi)$ within the spatially diffuse interface region. Simple forms of $q(\phi)$ are used to reproduce the desired $V$-dependent forms of $k(V)$ and $m(V)$ across an extremely wide velocity range, from near [$k(V) \to k_e$ where $k_e$ is the equilibrium value of the partition coefficient] to far from [$k(V) \to 1$] equilibrium conditions. 2D simulations revealed a new burgeoning instability in dendrite tip growth, driven by solute trapping at velocities nearing the absolute stability limit. These simulations also reproduced the formation of the widely observed banded microstructures and predicated band spacings that agree quantitatively with observations in rapidly solidified Al-Cu thin films \cite{mckeown_time-resolved_2016}, as shown in Fig.~\ref{fig:illustration}(b)-(c). Furthermore, the bands in the PF simulation form by the same lateral spreading mechanism as observed in the experiment, with a lateral velocity that remarkably matches the experimental measurements \cite{ji2024microstructure}. Furthermore, this PF model has recently been used to study the formation of banded microstructures during laser powder-bed fusion of a Magnesium alloy \cite{tourret2024emergence}.
The strategy of enhancing solute diffusivity to upscale interface thickness has also been implemented in a recent formulation \cite{li4989926phase} with finite interface dissipation \cite{steinbach_phase-field_2012} to simulate banded microstructures during rapid solidification.

In this paper, we revisit the aforementioned quantitative PF model \cite{ji_microstructural_2023}, hereafter referred to as Model I, and present a derivative Model II based on a similar variational framework. We derive the evolution equations of each model from a free-energy functional and perform the 1D asymptotic analysis, demonstrating that the solute trapping properties of both PF models match the predictions of the microscopic Continuous Growth (CG) model \cite{aziz_model_1982} in the large $V$ limit. While Model I is derived from a specific phenomenological free-energy functional introduced in Ref.~\cite{karma_phase-field_2003} for a dilute binary alloy, Model II is based on a free energy that interpolates between bulk phases at the interface, allowing for the modeling of both dilute and non-dilute alloys by integrating more general free-energy function fits derived from sources such as the Calculation of Phase Diagrams (CALPHAD).
In addition, we introduce a revised PF model that separates solute diffusion in normal and tangential directions to study the effects of excess surface diffusion. Furthermore, we expand the modeling capabilities by coupling latent-heat diffusion to the PF simulation and extending the simulation to 3D. 

Numerical simulations in 2D produce microstructure selection maps for dilute Al-Cu alloys under different thermal conditions, including the standard frozen temperature approximation assuming a fixed homogeneous temperature gradient and the time-dependent calculation of thermal diffusion within an adiabatic zone. In the latter, the banded microstructure is selected at a higher composition, which is in good agreement with the results of linear stability analyses \cite{karma_dynamics_1992,karma_interface_1993}, demonstrating the significant effects of latent-heat diffusion on banding. The transitions from cellular/dendritic to planar/banding with various alloy compositions are investigated through 2D PF simulations. The PF results with $S=1$ agree well with the absolute stability limit that is analytically predicted when the cellular/dendritic solution loses stability. The PF model is also extended to 3D, with both axisymmetric and full 3D simulations conducted at velocities close to the absolute stability limit. The latter reveal a tail instability developing in the grooves of dendrite arms, which resembles the corrugated roofs observed in 2D simulations but differs from the hemispherical caps seen in axisymmetric cases.

Moreover, we address a numerical issue specific to Model I under certain combinations of alloy and modeling parameters. 
The free-energy density $f_{AB}(\phi,T)$ in Model I encompasses contributions from both the pure substance and the solute addition, where $\phi$ is the PF varying from $+1$ in the solid to $-1$ in the liquid, and $T$ denotes temperature. 
A phenomenological form of $f_{AB}(\phi,T)$ \cite{karma_phase-field_2003} is chosen in Model I, such that the stationary solution of the PF profile is exactly a hyperbolic tangent function. This is achieved by using different functions of $\phi$ to interpolate the pure-substance and chemical parts of the free energy between the solid and liquid phases. 
In the dilute limit, this PF formulation is equivalent to the Kim-Kim-Suzuki model \cite{kim_phase-field_1999} that interpolates the concentration in the interface region between its solid and liquid values, and a thermodynamically-based formulation that interpolates the grand potential \cite{plapp_unified_2011}. 
The choice of $f_{AB}(\phi,T)$ in Model I, however, can lead to the emergence of an additional spurious minimum of the bulk free-energy density at values other than $\phi = \pm 1$ within the interval $-1<\phi<1$ , thereby causing unphysical simulation results. 
As presented in Sec.~\ref{Sec:Model_A}, we distinguish between the``safe'' and ``unsafe'' regions within the parameter space. In the safe region, Model I operates under any interface driving force, whereas in the unsafe region, certain driving forces might induce the development of unphysical phases. Typically, the unsafe region is characterized by a high nominal concentration $c_{\infty}$, a small $k_e$, and/or a large interface thickness $S$. Even for the case of $S=1$, a sufficiently small $k_e$ can lead to the appearance of unphysical phases under certain driving forces, which holds true for both Model I, as discussed here, and the PF models in Refs.~\cite{kim_phase-field_1999, plapp_unified_2011}. Within the unsafe region, simulations using these models are still feasible as long as the driving force does not trigger unphysical phases, though extra caution is required.

This numerical issue does not arise in Model II, which interpolates the free energy of the solid and liquid phases across a diffuse interface. It should be noted that Model II in its $S=1$ case is equivalent to the PF formulation introduced in Refs.~\cite{wheeler_phase-field_1992,warren_prediction_1995}, where the interface is assumed to be in an intermediate state between the two bulk phases. Although the solution of Model II deviates from the hyperbolic tangent profile of the stationary PF solution, this model reproduces the quantitative $k(V)$ and $m(V)$ curves similar to that of Model I for dilute alloys and circumvents the problem of unphysical phases for any combination of alloy and modeling parameters. However, given that Model II introduces a higher energy barrier, it requires more stringent conditions for numerical stability and results in higher computational costs. 

Thus, we primarily utilize Model I for PF simulations of dilute alloys, resorting to Model II only when parameters fall into the unsafe region. The approach of interpolating the free energy of bulk phases at the interface in Model II also provides additional flexibility. This allows for the integration of Gibbs free energy as functions of temperature and alloy concentration for each phase, commonly used to accurately predict phase properties in CALPHAD. By applying these experimentally verified free-energy functions and enhancing interfacial diffusion for an upscaled interface, we investigate solute trapping for concentrated Al-Ag alloys. The nonequilibrium liquidus and solidus predicted by the PF model are nearly $S$-independent, enabling the quantitative modeling of rapid solidification of non-dilute binary alloys at experimentally relevant time and length scales. An example 2D PF simulation for a concentrated Al-Ag alloy is provided.

The paper is organized as follows. In Sec.~\ref{Sec:Sharp_interface}, we first introduce the sharp-interface equations for a moving solid-liquid interface. Then, we present detailed derivations of PF models in Sec.~\ref{Sec:phase_field_model}. Asymptotic analyses, approximate and full PF solutions of solute trapping are presented in Sec.~\ref{Sec:Asymptotic_and_PF_solutions}. Numerical results in both 2D and 3D are discussed in Sec.~\ref{Sec:Numerical_results}. Lastly, conclusions and perspectives are given in Sec.~\ref{Sec:Conclusions_perspectives}.

\section{Sharp-interface equations} \label{Sec:Sharp_interface}

Consider the solidification of a dilute binary alloy composed of substances A and B, where the liquidus and solidus lines in the phase diagram are approximated as straight lines with slope magnitudes $m>0$ and $m/k>0$, respectively. The partition coefficient $k$ is defined as the ratio $c_s/c_l$, where $c_s$ and $c_l$ represent the concentrations (in molar fractions) of impurity B on the solid and liquid sides of the interface, respectively. For a solid-liquid interface moving at a velocity $V$, the interface temperature satisfies the generalized Gibbs-Thomson relation:
\begin{equation}
T(V)=T_M-m(V) c_l-\Gamma \mathcal{K}-V / \mu_k, \label{Gibbs_Thomson}
\end{equation}
where $T_M$ is the melting temperature of pure A, $\mathcal{K}$ is the interface curvature, and $\mu_k$ is the linear kinetic coefficient. The Gibbs-Thomson constant $\Gamma$ is defined as
\begin{equation}
\Gamma=\frac{\gamma T_M}{\Delta h_f},
\end{equation}
where $\gamma$ is the surface tension and $\Delta h_f$ is the latent heat of fusion per unit volume. For simplicity, both the surface tension and the kinetic coefficient are assumed to be isotropic here, and the consideration of anisotropic interface properties will be addressed in subsequent sections.

During rapid solidification, the solid-liquid interface deviates from thermodynamic equilibrium. This deviation results in the trapping of excess solute as the diffuse interface region crystallizes before the completion of solute partitioning. Consequently, both the partition coefficient and the liquidus slope exhibit a nonlinear dependence on the continuous interface velocity $V$, transitioning from their equilibrium values $k_e$ and $m_e$ to their maximum values under complete solute trapping. The microscopic CG model provides a theoretical description of this phenomenon \cite{aziz_model_1982,aziz_continuous_1988,aziz_transition_1994}, which assumes standard diffusion combined with attachment-limited kinetics at a sharp solid-liquid interface of ideal binary alloys. This model predicts steady-state solutions for solute trapping properties of a planar interface. The nonequilibrium partition coefficient is given by:
\begin{equation}
k(V)=\frac{k_e+V/V_d}{1+V/V_d}, \label{CGM_k}
\end{equation}
where $V_d$ is the so-called diffusive velocity that generally depends on alloys. According to Eq.~\eqref{CGM_k}, the nonequilibrium partition coefficient varies from $k(V) \to k_e$ at $V \ll V_d$ toward $k(V) \to 1$ at $V \gg V_d$. 
Meanwhile, the nonequilibrium liquidus slope in the dilute solution limit of the CG model has the form:
\begin{equation}
\frac{m(V)}{m_e}=\frac{1 - k(V) + [k(V) + (1 - k(V))\alpha]\ln[k(V)/k_e] }{1-k_e}, \label{m_me}
\end{equation}
where $k(V)$ follows Eq.~\eqref{CGM_k}. The coefficient $\alpha$ indicates the extent to which the interface is ``dragged'' due to the presence of solute atoms: a value of 1 stands for full solute drag, while 0 indicates negligible solute drag. According to Eq.~\eqref{m_me}, the nonequilibrium liquidus slope varies from $m(V) \to m_e$ at $V \ll V_d$ toward $m(V) \to m_e{ \ln (1 / k_{e})}/{(1-k_e)}$ at $V \gg V_d$. 

In PF models with a diffuse solid-liquid interface of finite thickness, solute trapping properties can be effectively reproduced, as shown in Fig.~\ref{fig:q_phi}. As will be discussed in Sec.~\ref{Sec:PF_solutions_1D}, the quantitative PF model with an upscaled interface thickness matches Eq.~\eqref{Gibbs_Thomson} that describes the interface capillarity and kinetics when the proper coupling constants are chosen. In addition, the PF model reproduces $k(V)$ and $m(V)$ functions from Eqs.\ \eqref{CGM_k} and \eqref{m_me} in asymptotic analyses assuming a hyperbolic tangent PF profile and a large $V$ limit, as will be presented in Sec.~\ref{Sec:Asymptotic}.

\section{Quantitative phase-field models} \label{Sec:phase_field_model}
In this section, we first derive generic variational PF models for the solidification of binary alloys, including Model I for dilute alloys in Sec.~\ref{Sec:Model_A}, and Model II for both dilute and non-dilute alloys in Sec.~\ref{Sec:Model_B}. Then, we revise Model I to cancel the excess surface diffusion in Sec.~\ref{Sec:surface_diffusion}. Different thermal conditions are discussed in Sec.~\ref{Sec:Thermal_conditions}.

\subsection{Model I} \label{Sec:Model_A}

The total free energy of a two-phase system for alloy solidification can be expressed as \cite{karma_phase-field_2003}
\begin{equation}
F[\phi, c, T]=\int_{d V}\left[\frac{\sigma}{2}|\vec{\nabla} \phi|^{2}+f_{A B}(\phi, c, T)\right]. \label{F_functional}
\end{equation}
Within the integrand, the first term represents the gradient energy, ensuring a finite interface thickness. The second term, $f_{A B}(\phi, c, T)$, is the bulk free-energy density of a binary mixture of A and B atoms/molecules, with $c$ denoting the solute concentration, defined as the mole fraction of B. In the dilute limit, the free-energy density is given by
\begin{align}
f_{A B}(\phi, c, T)=&f\left(\phi, T_{M}\right)+f_{T}(\phi) (T-T_M) \nonumber \\
&+\frac{R T_{M}}{v_{0}}(c \ln c-c)+\epsilon(\phi) c, \label{f_AB}
\end{align}
where $f\left(\phi, T_{M}\right)$ is a double-well potential providing the stability of the two bulk phases $\phi=\pm1$. We choose a standard form:
\begin{equation}
f\left(\phi, T_M\right) = h\left(-\frac{\phi^{2}}{2}+\frac{\phi^{4}}{4}\right),
\end{equation}
with $h$ representing the energy barrier height between local minima corresponding to the solid and liquid. In Eq.~\eqref{f_AB}, the free-energy density combines the contribution from solute addition and that of the pure material, where the latter is denoted by $f(\phi, T)$ and is expanded to the first order in $(T-T_M)$ with $f_{T}(\phi) \equiv \partial f(\phi, T) /\partial T |_{T=T_{M}}$. The term $R T_M v_{0}^{-1}(c \ln c-c)$ is the standard entropy of mixing, where $R$ is the gas constant and $v_0$ is the molar volume assumed to be constant. The term $\epsilon(\phi) c$ is the enthalpy of mixing, where
\begin{equation}
\epsilon(\phi)=\frac{1+g(\phi)}{2} \epsilon_{s}+\frac{1-g(\phi)}{2} \epsilon_{l},    
\end{equation}
interpolating between the values $\epsilon_s$ and $\epsilon_l<\epsilon_s$ in the solid and liquid, respectively. Thus, it can also be expressed as $\epsilon(\phi)=[\bar{\epsilon}+g(\phi)\Delta \epsilon/2]$, where $\bar{\epsilon} \equiv(\epsilon_{s}+\epsilon_{l}) / 2$ and $\Delta \epsilon=(\epsilon_{s}-\epsilon_{l})$. Here $g(\phi)$ is an interpolating function that satisfies $g^{\prime}( \pm 1)=g^{\prime \prime}( \pm 1)=0$ and guarantees that the local minima of the free-energy density remain at $\phi=\pm 1$. We use a quantic form of $g(\phi)$ that satisfies the constraints:
\begin{equation}
g(\phi)=\frac{15}{8} \left( \phi- \frac{2}{3} \phi^{3}+ \frac{1}{5} \phi^{5} \right). \label{g_phi}
\end{equation}

The system is expected to relax to a global free-energy minimum. The dynamical evolution of the scalar fields $c$ and $\phi$ follows the standard variational forms for conserved and non-conserved dynamics, respectively:
\begin{align}
\frac{\partial c}{\partial t}&=\vec{\nabla} \cdot\left(K_{c} \vec{\nabla} \frac{\delta F}{\delta c}\right), \label{functional_dev_c} \\
\frac{\partial \phi}{\partial t}&=-K_{\phi} \frac{\delta F}{\delta \phi}, \label{functional_dev_p}
\end{align}
where $K_{c}$ represents the mobility of solute atoms or molecules, and $K_{\phi}$ is a coefficient associated with the interface kinetics. In equilibrium, $\partial_t c=\partial_t \phi=0$, and Eqs.\ \eqref{functional_dev_c}-\eqref{functional_dev_p} simplify to ${\delta F}/{\delta c}=\mu_{E}^{c}$ and ${\delta F}/{\delta \phi}=0$, where $\mu_{E}^{c}$ is the spatially uniform equilibrium value of the chemical potential. Applying the first equilibrium condition to a planar interface, we derive the stationary concentration profile
\begin{equation}
c_0(x)=c_l^0\exp(b\left[1+g(\phi_0(x))\right]), \label{c_0_x}
\end{equation}
where $\phi_0(x)$ is the stationary PF profile, $b =\ln k_e/2<0$, $k_e \equiv c_s^0/c_l^0 =\exp \left(-{v_{0} \Delta \epsilon}/{R T_{M}}\right)$, and $c_s^0$ and $c_l^0$ are the equilibrium concentrations at the solid and liquid sides of the interface, respectively. Applying the second equilibrium condition, one can show that $\phi_0$ follows a hyperbolic tangent profile of the form
\begin{equation}
\phi_0(x)=-\tanh \left( \frac{x}{\sqrt{2} W} \right), \label{phi_0_x}
\end{equation}
leading to the function 
\begin{equation}
f_T(\phi)=-\frac{R T_M}{v_0 m_e} \exp (b[1+g(\phi)]) \label{f_T_phi}
\end{equation}
that reproduces the relation $T=(T_M-m_e c_l^0)$. To complete the model, we choose $K_c=v_{0} D_l q(\phi) c /(R T_{M})$, ensuring that Eq.~\eqref{functional_dev_c} adheres to Fick's law of diffusion in the bulk phase, where $q(\phi)$ is a function that interpolates solute diffusivity between $D_l$ in the liquid and $D_s$ in the solid. Consequently, we obtain the evolution equations
\begin{align}
\tau_0 \frac{\partial \phi}{\partial t}=&W^{2} \nabla^{2} \phi + \phi -\phi^3 \label{model_A_phi} \\
&-\lambda g^{\prime}(\phi)\left[c+\frac{(T-T_M)}{m_e} \exp(b[1+g(\phi)])\right], \nonumber \\
\frac{\partial c}{\partial t}=& \vec{\nabla} \cdot \left\{D_l q(\phi) c \vec{\nabla}[\ln c-b g(\phi)] \right\}, \label{model_A_c}
\end{align}
where we define $\tau_0=1 /\left(K_{\phi} h\right)$, $W=(\sigma / h)^{1 / 2}$, and $\lambda \equiv-b R T_{M} /\left(v_{0} h\right)>0$. 

As discussed later in Sec.~\ref{Sec:PF_1D_dilute}, the connection between PF and material parameters is established through the relations
\begin{equation}
\tau_0 = \frac{(SW_0)^2}{\Gamma_0 \mu_k^0} \label{tau0}
\end{equation}
with $\Gamma=\Gamma_0$ and $\mu_k=\mu_k^0$ for an isotropic interface, and 
\begin{equation}
\lambda=a_{1}^0 SW_0 \frac{b}{ (k_e-1)} \frac{m_e}{\Gamma_0}, \label{lambda}
\end{equation}
where 
\begin{equation}
a_1^0=SW_0\int_{-\infty}^{\infty} dx \left[ \frac{d\phi_0(x)}{dx} \right]^2
\end{equation}
is a constant linked to the stationary PF profile. With the hyperbolic tangent PF profile in Eq.~\eqref{phi_0_x}, we obtain $a_1^0 = 2\sqrt{2}/3$.

\begin{figure}[htbp!]
\includegraphics[scale=1]{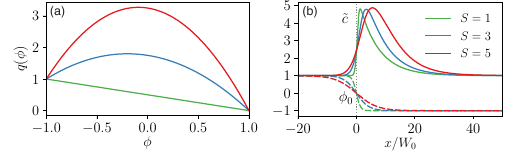}
\centering
\caption{
(a) Plots of $q(\phi)$ for $S=1,3,5$ ($A=1,6,12$), with (b) corresponding PF $\phi=\phi_0$ and normalized concentration $\tilde{c}=c/c_\infty$ profiles obtained from the numerical solution of Eq.~\eqref{c_scale}.
\label{fig:q_phi}
}
\end{figure}

For the one-sided model of alloy solidification with $D_s = 0$, $q(\phi)=(1-\phi)/2$ is the simplest form that describes the anticipated monotonous decrease in diffusivity from liquid to solid across the interface. However, this form leads to spurious excess trapping at a lower $V$ when $S \gg 1$, given that the diffusive speed in the PF model is $V_d = V_d^0/S$, with $V_d^0\equiv D_l/W_0$ being a constant. To eliminate this excess trapping, we adopt the quadratic form 
\begin{equation}
q(\phi)=A\frac{(1-\phi)}{2}-(A-1)\frac{(1-\phi)^2}{4}, \label{q_phi}
\end{equation}
which enhances $D(\phi)$ within the interface region. Here, the coefficient $A \ge 1$ is an adjustable parameter for different $S$ values, with a larger $A$ corresponding to a larger solute diffusivity in the interface region. As discussed in \cite{ji_microstructural_2023}, for a specific $S$ value, the interfacial diffusivity coefficient $A$ is determined by minimizing discrepancies with reference $k(V)$ and $m(V)$ curves that correspond to $S=1$ and $A=1$ across a relevant velocity range. This procedure should be repeated for different alloy systems to determine the optimal values of $A$. For dilute Al-Cu alloys, the choices of $A=6$ for $S=3$ and $A=12$ for $S=5$ reproduce the same trapping properties as the reference model with $A=1$ and $S=1$ \cite{ji_microstructural_2023}. For simplicity, we use the same $A$ values for all PF simulations in this paper.

\subsubsection*{Limitations of Model I}

The phenomenological free-energy density in Eq.~\eqref{f_AB} has contributions from both the pure substance and the solute addition. The choice of the exponential form of $f_T(\phi)$ in Eq.~\eqref{f_T_phi} guarantees a standard hyperbolic tangent PF profile as its stationary solution. However, under specific combinations of alloy and modeling parameters, this formulation introduces an additional spurious minimum of the free-energy density at intermediate values of $\phi$ between the $\phi = \pm 1$ minima of the bulk solid and liquid. This spurious minimum, which corresponds to a stable spatially uniform fixed point of the $\phi$ dynamics, can potentially lead to the development of an unphysical phase corresponding to this spurious minimum in PF simulations. Considering the equivalence between Model I with $S$ = 1 and the PF models introduced in Refs.~\cite{kim_phase-field_1999,plapp_unified_2011} in the dilute limit, it is anticipated that other PF models will also face the same fixed-point issue.

To illustrate this issue, we write down the evolution equation for $\phi$ under isothermal conditions in the high velocity limit of complete trapping (also commonly known as massive transformation) where the concentration is spatially uniform
$c=c_{\infty}$. In this limit, Eq. (\ref{model_A_phi}) neglecting the Laplacian term simplifies to 
\begin{equation}
\tau_0\frac{\partial \phi}{\partial t}  = \bar{f}_{\phi}(\phi),
\label{phidyn}
\end{equation}
where the function $\bar{f}_{\phi}(\phi)\sim - \partial f_{A B}(\phi,c_\infty,T)/\partial \phi$ is given by
\begin{equation}
\bar{f}_{\phi}(\phi)=\phi-\phi^3-\tilde{\lambda} g^{\prime}(\phi)\left(1+ \widetilde{T} \, e^{b[1+g(\phi)]}\right),
\label{eq:fixed_point}
\end{equation}
with $\tilde{\lambda} \equiv \lambda c_{\infty}$ and $\widetilde{T}=(T-T_M)/(m_e c_{\infty})$ being dimensionless coefficients. Our focus here is on identifying the parameter space within which the simulations remain ``safe'', meaning without an additional stable fixed point of the $\phi$ dynamics governed by Eq.~\eqref{phidyn} at an intermediate value of $\phi$ between $-1$ and $1$ corresponding to a spurious minimum of $f_{AB}(\phi,c_\infty,T)$. In the ``unsafe'' region, there always exists a $\widetilde{T}$ value that produces additional fixed points between $\phi = \pm 1$.
The boundary separating these two regions is determined by the conditions:
\begin{equation}
\bar{f}_{\phi}(\phi)= 0,\quad \frac{\partial \bar{f}_{\phi}(\phi)}{\partial \phi}= 0,\quad \frac{\partial^2 \bar{f}_{\phi}(\phi)}{\partial \phi^2}= 0.
\label{eq:saddle}
\end{equation}

\begin{figure}[htbp!]
\includegraphics[scale=0.67]{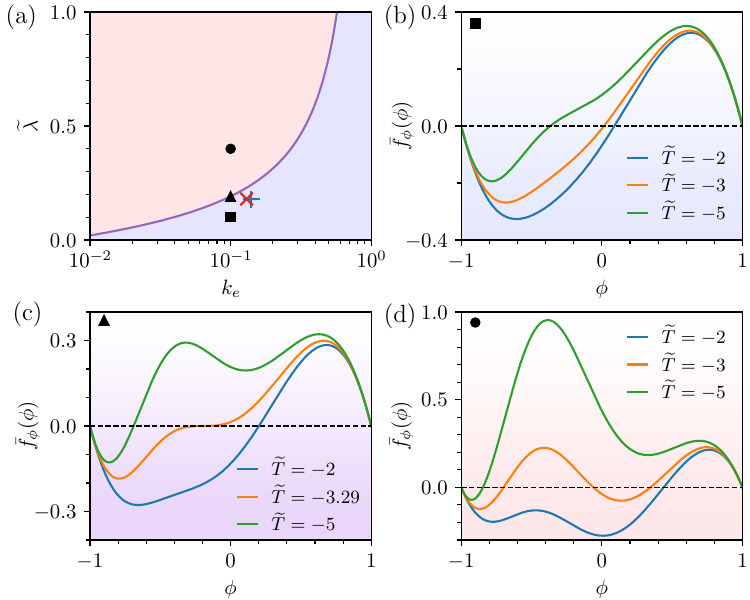}
\centering
\caption{
Stability analysis of Model I. (a) Parameter space of $\tilde{\lambda}$ versus $k_e$. The model is stable for parameters within the light blue region and unstable within the light red region. The blue plus symbol and red cross symbol represent Al-3wt.\%Cu and Al-1wt.\%Si, respectively. Representative $\bar{f}_{\phi} (\phi)$ curves at different $\widetilde{T}$ values are shown for the safe region in (b), at the boundary line in (c), and in the unstable region in (d).
\label{fig:fixed_point}
}
\end{figure}

We distinguish the safe and unsafe regions in the parameter space of $\tilde{\lambda}$ versus $k_e$, as shown in Fig.~\ref{fig:fixed_point}(a). At the point $k_e = 0.1$ and $\tilde{\lambda} = 0.1$ within the safe region, Fig.~\ref{fig:fixed_point}(b) illustrates the function $\bar{f}_{\phi}(\phi)$ for different $\widetilde{T}$ values, showing that fixed points of $\phi$ exist only at the two physical bulk phases $\phi = \pm 1$, where $\bar{f}_{\phi}(\phi)=0$ and $\bar{f}_{\phi}^{\prime}(\phi)$ is negative. As $\tilde{\lambda}$ increases to approximately 0.19, the system approaches the boundary of the safe region, where $\widetilde{T} = -3.29$ yields a saddle point near $\phi = -0.2$, as shown by the orange line in Fig.~\ref{fig:fixed_point}(c). At the point $k_e = 0.1$ and $\tilde{\lambda} = 0.4$ within the unsafe region, $\widetilde{T} = -3$ produces two additional fixed points where $\bar{f}_{\phi}(\phi)=0$ as shown by the orange line in Fig.~\ref{fig:fixed_point}(d). Those additional fixed points are stable (unstable) where $\bar{f}_{\phi}^{\prime}(\phi)$ is negative (positive). Consequently, an unphysical phase corresponding to the additional stable fixed point may suddently appear in PF simulations at this driving force. However, the presence of this stable fixed point is only a necessary but not sufficient condition for this phase to appear since Eq. (\ref{phidyn}) neglects the contribution of the gradient square term in the free-energy. Moreover, for sufficiently small ($\widetilde{T} = -2$) or large ($\widetilde{T} = -5$) driving forces, no additional fixed points are observed. 

It should be noted that even within the unsafe region, simulations using Model I are feasible as long as the interface driving force does not reach a value that would induce additional fixed points, such as the simulations of Al-9wt.\%Cu in Ref.~\cite{ji_microstructural_2023}. Such simulations are possible when driving forces are relatively small, as illustrated by the blue line in Fig.~\ref{fig:fixed_point}(d).

\subsection{Model II} \label{Sec:Model_B}

The free-energy density of a binary alloy given by Eq.~\eqref{f_AB} is only applicable in the dilute limit. For a more general form of $f_{AB}$, the binary mixture part can be expressed as an interpolation between two bulk phases:
\begin{align}
f_{A B}(\phi, c, T) & \nonumber \\
= f\left( \phi, T_M \right)+&\frac{1+g(\phi)}{2}f_s(c,T)+\frac{1-g(\phi)}{2}f_l(c,T), \label{f_AB_general}
\end{align}
where $f_s(c,T)$ and $f_l(c,T)$ are the free energy densities of the solid and liquid phases, respectively. These densities are related to the Gibbs free energies, typically obtained from CALPHAD, through the relations $f_s(c,T) = {G_s(c,T)}/{v_0}$ and $f_l(c,T) = {G_l(c,T)}/{v_0}$. The PF model based on Eq.~\eqref{f_AB_general}, referred to as Model II, is applicable for modeling both dilute and non-dilute binary alloys. Using the same quartic form of $g(\phi)$ as in Eq.~\eqref{g_phi}, Model II avoids the formation of additional fixed points for any combination of modeling parameters or driving forces, which can be readily demonstrated by following the same stability analysis presented in Sec.~\ref{Sec:Model_A}.
Below, we first derive Model II in the dilute limit, and then we discuss its application to general binary alloys with the capability of integrating complex free-energy functions for each phase.

\subsubsection{Dilute binary alloys}

In the dilute limit, the free-energy densities of the solid and liquid phases take the forms
\begin{equation}
f_s(c,T)=-\frac{R T_M}{v_0 m_e} k_e (T-T_M) + \frac{R T_M}{v_0}(c \ln c-c)+\epsilon_s c \label{fs_dilute}
\end{equation}
and
\begin{equation}
f_l(c,T)=-\frac{R T_M}{v_0 m_e} (T-T_M) + \frac{R T_M}{v_0}(c \ln c-c)+\epsilon_l c, \label{fl_dilute}
\end{equation}
respectively. Substituting Eqs.~\eqref{fs_dilute}-\eqref{fl_dilute} into Eq.~\eqref{f_AB_general}, the expression for $f_{AB}(\phi, c, T)$ simplifies to the same form as Eq.~\eqref{f_AB}, with
\begin{equation}
f_T(\phi)=-\frac{R T_M}{v_0 m_e} \frac{1}{2}\left[1+k_e-(1-k_e)g(\phi) \right]. \label{f_T_B}
\end{equation}
This form of $f_T(\phi)$ effectively interpolates between the same bulk values $f_T(+1)$ and $f_T(-1)$ as those of Model I, as given in Eq.~\eqref{f_T_phi}. While Model I ensures a hyperbolic tangent function for the stationary PF profile $\phi_0$, the choice of $f_T(\phi)$ in Eq.~\eqref{f_T_B} for Model II does not lead to the same stationary solution.

Following the variational dynamics presented in Sec.~\ref{Sec:Model_A}, the evolution equation of $\phi$ in Model II for a dilute alloy is:
\begin{align}
\tau_0 \frac{\partial \phi}{\partial t}=&W^{2} \nabla^{2} \phi+\phi-\phi^{3} \nonumber \\
&-\lambda g^{\prime}(\phi)\left[c+\frac{(T-T_M)}{m_e} \frac{(k_e-1)}{2b} \right]. \label{phi_newfT}
\end{align}
Sine the evolution of the concentration field remains the same as Eq.~\eqref{model_A_c}, we continue to use the same quadratic form for $ q(\phi) $ in Eq.~\eqref{q_phi} and the coefficient $ A $ for different $ S $ values.
As demonstrated in Sec.~\ref{Sec:PF_1D_dilute}, Model II reproduce quantitatively the $k(V)$ and $m(V)$ curves similar to that of Model I for a dilute alloy, and its dynamical solution converges to a hyperbolic tangent PF profile at high $V$.
Furthermore, unlike Model I, Model II does not suffer from the formation of spurious phases. This is because the form of Eq. (\ref{phi_newfT}) implies that $\phi$ has only two stable spatially uniform fixed points at $\phi=\pm 1$ for arbitrarily large value of the driving force corresponding to the term in square bracket on the right-hand-side of Eq. (\ref{phi_newfT}). This is easily seen by setting $\partial_t\phi=0$, which yields a quartic equation in $\phi$ that has four real roots, two stable ones at $\phi=\pm 1$ and two unstable ones at $\phi<-1$ and $-1<\phi<1$. The same holds true for the case of a general binary alloy discussed next.

\subsubsection{General binary alloys} \label{Sec:ModelB_general}

Model II can be generalized to accommodate both dilute and non-dilute binary alloys, which takes into account the complex free-energy functions that usually result in nonlinear liquidus and solidus. This generalization allows for a more accurate prediction of the solidification behavior in alloy systems such as hypo-eutectic Al-Ag alloys. To derive the generalized Model II, we start from Eq.~\eqref{f_AB_general} and follow the dynamics described in Eqs.~\eqref{functional_dev_c} and \eqref{functional_dev_p} to obtain the evolution equations:
\begin{align}
\tau_0 \frac{\partial \phi}{\partial t} =& W^2 \frac{\partial^2 \phi}{\partial x^2} +\phi-\phi^3 -\frac{1}{2h} g^{\prime}(\phi) f_s(c,T) \nonumber \\ 
&+\frac{1}{2h} g^{\prime}(\phi) f_l(c,T). \label{dpdt_B_general}
\end{align}
and
\begin{align}
\frac{\partial c}{\partial t}=\vec{\nabla} \cdot&\left[ D_l q(\phi) c (1-c) \vec{\nabla}\left(\frac{1+g(\phi)}{2 h_0} \partial_c f_s(c,T) \right. \right. \nonumber \\
&\left.\left.+\frac{1-g(\phi)}{2 h_0} \partial_c f_l(c,T)\right)\right], \label{dcdt_B_general}
\end{align}
where $h_0 \equiv R T_M / v_0$. Similar to Model I and Model II in its dilute limit, we choose $\tau_0=1 /\left(K_\phi h\right)$ and $ W=(\sigma / h)^{1 / 2}$. However, we set $K_c=v_0 D_l q(\phi) c (1-c) /\left(R T_M\right)$ to restore the Fickian diffusion in non-dilute alloys.

To implement this model, explicit forms of $f_s(c,T)$ and $f_l(c,T)$ based on the equilibrium thermodynamic properties of the two-phase system are required, typically obtained from CALPHAD. The relation between the relaxation time $\tau_0$ and the physical parameters remains consistent with Eq.~\eqref{tau0}, as in the case for a pure substance \cite{karma_quantitative_1998}. Given that the barrier height $h$ appears explicitly in the evolution equation, we determine its value using the relation for a pure substance $h = \gamma_0 / (a_1^0 W)$, where the surface tension $\gamma_0$ is related to the Gibbs-Thomson coefficient by $\Gamma_0 = {\gamma_0 T_M} / {\Delta h_f}$. With known free-energy functions, the latent heat of fusion per unit volume can be derived in the pure substance limit through the relation
\begin{equation}
\Delta h_f = T_M \left.\left[\partial_T f_s(c,T) - \partial_T f_l(c,T) \right]\right|_{T=T_M,c\to0}. \label{Delta_h_f}
\end{equation}
Thus, an additional input for $\gamma_0$ is not necessary once $\Gamma_0$ and free-energy functions are specified in a PF simulation. 

Using Gibbs free energy data \cite{witusiewicz_agcu_2004} for the solid $G_s(c,T)$ and liquid $G_l(c,T)$ phases of the Al-Ag alloy, both in units of J/mol, we have 
\begin{widetext}
\begin{align}
& v_0  f_l(c, T) = R T \left[(1 - c) \ln(1 - c) + c \ln c \right] + (1 - c) \, { }^{\circ} G_{\mathrm{Al}}^{\mathrm{Liq}}(T) + c \, { }^{\circ} G_{\mathrm{Ag}}^{\mathrm{Liq}}(T) + c (1 - c) \Big[{ }^0 L_{\mathrm{Al}, \mathrm{Ag}}^{\mathrm{Liq}}(T) \label{fl_Al_Ag} \\
&\, - { }^1 L_{\mathrm{Al}, \mathrm{Ag}}^{\mathrm{Liq}}(T) \, (2(1 - c) - 1) + { }^2 L_{\mathrm{Al}, \mathrm{Ag}}^{\mathrm{Liq}}(T) \, (2(1 - c) - 1)^2 - { }^3 L_{\mathrm{Al}, \mathrm{Ag}}^{\mathrm{Liq}}(T) \, (2(1 - c) - 1)^3 + { }^4 L_{\mathrm{Al}, \mathrm{Ag}}^{\mathrm{Liq}}(T) \, (2(1 - c) - 1)^4 \Big], \nonumber \\
& v_0 f_s(c, T) = R T \left[(1 - c) \ln(1 - c) + c \ln c \right] + (1 - c) \, G H_{\mathrm{Al}}^{\mathrm{SER}}(T) + c \, G H_{\mathrm{Ag}}^{\mathrm{SER}}(T) \label{fs_Al_Ag} \\
&\, + c (1 - c) \Big[{ }^0 L_{\mathrm{Al}, \mathrm{Ag}}^{\mathrm{fcc}}(T) - { }^1 L_{\mathrm{Al}, \mathrm{Ag}}^{\mathrm{fcc}}(T) \, (2(1 - c) - 1) + { }^2 L_{\mathrm{Al}, \mathrm{Ag}}^{\mathrm{fcc}}(T) \, (2(1 - c) - 1)^2 - { }^3 L_{\mathrm{Al}, \mathrm{Ag}}^{\mathrm{fcc}}(T) \, (2(1 - c) - 1)^3 \Big]. \nonumber
\end{align}
\end{widetext}
Further details on the thermodynamic functions of $T$ in Eqs.~\eqref{fl_Al_Ag}-\eqref{fs_Al_Ag} can be found in Eq.~(4) and Tables 3 and 4 of Ref.~\cite{witusiewicz_agcu_2004}. Based on these inputs and Table \ref{tab:table1}, we obtain $\Delta h_f = 10^9 \, \mathrm{J/m^3}$ according to Eq.~\eqref{Delta_h_f}, and $\gamma_0 = 0.21 \, \mathrm{J/m^2}$. 
The 1D solutions of this model across various solute concentrations will be discussed in Sec.~\ref{Sec:PF_1D_Al_Ag}, and an example 2D PF simulation for a concentrated Al-Ag alloy is presented in Sec.~\ref{Sec:PF_2D_Al_Ag}. For these simulations, we also use $R = 8.31 \, \mathrm{J/(mol \, K)}$ and $v_0 = 1.068 \times 10^{-5} \, \mathrm{m^3/mol}$.

\subsection{Model without excess surface diffusion} \label{Sec:surface_diffusion}

The enhancement of solute diffusivity in the interfacial region counteracts excess trapping perpendicular to the interface but also amplifies surface diffusion tangential to an upscaled interface. This phenomenon occurs in both models I and II, and across both dilute and non-dilute alloys. For simplicity, the following discussion focuses only on a dilute alloy using Model I, but the same methodology is applicable to other scenarios. 

To assess the effect of enhanced surface diffusion, we adapt the dynamics of solute concentration in Eq.~\eqref{functional_dev_c} to distinguish between solute diffusion components normal and tangential to the interface
\begin{equation}
\frac{\partial c}{\partial t}=\vec{\nabla} \cdot\left[K_{\bot}(\phi,c) \vec{\nabla}_{\bot} \mu^c+K_{\parallel}(\phi,c) \vec{\nabla}_{\parallel} \mu^c \right], \label{diff_two}
\end{equation}
where $\vec{\nabla}_{\bot} \equiv \mathbf{n}(\mathbf{n} \cdot \vec{\nabla})$, $\vec{\nabla}_{\parallel} \equiv [\vec{\nabla}-\mathbf{n}(\mathbf{n} \cdot \vec{\nabla})]$, $\mathbf{n}=-\vec{\nabla} \phi / |\vec{\nabla} \phi|$ represents the unit vector normal to the interface pointing towards the liquid, and $\mu^c=\delta F / \delta c$ denotes the chemical potential. $K_{\bot}(\phi,c)$ and $K_{\parallel}(\phi,c)$ represent the mobility of solute atoms or molecules normal and tangential to the interface, respectively, and they have the forms
\begin{align}
K_{\bot}(\phi,c) &= v_{0} D_l q_{\bot}(\phi) c /(R T_{M}), \label{K_bot} \\
K_{\parallel}(\phi,c) &= v_{0} D_l q_{\parallel}(\phi) c /(R T_{M}). \label{K_parallel}
\end{align}
With the interpolation functions $q_{\bot}(\phi) = q_{\parallel}(\phi) = q(\phi)$ for $S = 1$, Eq.~\eqref{diff_two} simply reduces to Eq.~\eqref{functional_dev_c}. For $S>1$, in order to enhance the solute diffusion only normal to the interface, we choose $q_{\bot}(\phi) = q(\phi)$ with a quadratic form in Eq.~\eqref{q_phi}, and
\begin{equation}
q_{\parallel}(\phi)=\frac{1-\phi}{2}.
\end{equation}
Thus, while the normal component $q_{\bot}(\phi)$ is a nonlinear function that depends on $S$, $q_{\parallel}(\phi)$ remains a simple linear interpolation function for all $S$. From Eq.~\eqref{diff_two}, we derive
\begin{equation}
\frac{\partial c}{\partial t}=\vec{\nabla} \cdot\left(K_{\parallel} \vec{\nabla} \mu^c\right)+\vec{\nabla} \cdot\left[\left(K_{\bot}-K_{\parallel}\right) \frac{\vec{\nabla} \phi\left(\vec{\nabla} \phi \cdot  \vec{\nabla} \mu^c\right)}{|\vec{\nabla} \phi|^{2}}\right]. \label{c_t_K}
\end{equation}
Combining Eqs.\ \eqref{K_bot}, \eqref{K_parallel}, and \eqref{c_t_K}, we arrive at
\begin{align}
&\frac{\partial c}{\partial t}=D_l \vec{\nabla} \cdot \left \{ q_{\parallel}(\phi) c \vec{\nabla} [\ln c-b g(\phi)] \right \}+\\
&D_l \vec{\nabla} \cdot \left \{ [q_{\bot}(\phi)-q_{\parallel}(\phi)] c \frac{\vec{\nabla} \phi \left[\vec{\nabla} \phi \cdot \vec{\nabla} \left(\ln c-b g(\phi) \right)\right]}{|\vec{\nabla} \phi|^2} \right \}. \nonumber
\end{align}
The variational analysis in Appendix \ref{Appendix:Surf_diff} demonstrates that this formulation adheres to the gradient dynamics, ensuring that the free-energy functional $F$ decreases monotonically over time.

Given that we formulated this PF model on the basis of Model I, the PF evolution equation should match Eq.~\eqref{model_A_phi}. It is also worth noting that the PF solution of this model in 1D mirrors that of Model I since the solute diffusion perpendicular to the interface remains the same. Upon comparing dendrite shapes from 2D PF simulations in Sec.~\ref{Sec:2D_dendrite_shapes}, both with and without excess surface diffusion, it becomes evident that the effect of surface diffusion varies based on the choice of $S$.


\subsection{Thermal conditions} \label{Sec:Thermal_conditions}

In any of the PF equations given by \eqref{model_A_phi}, \eqref{phi_newfT}, or \eqref{dpdt_B_general}, the temperature $T$ can be treated as a constant under isothermal conditions. Meanwhile, other thermal conditions can be integrated into the PF simulations by making $T$ a function of time and space. These include the frozen temperature approximation (FTA), which assumes that thermal diffusion occurs much more rapidly than solute diffusion, thereby treating the temperature field as ``frozen'', and the thermal field calculation (TFC) that is directly coupled to the PF equation \cite{song_thermal-field_2018}.

\paragraph*{Frozen Temperature Approximation}

In the context of directional solidification, the sample is pulled by a speed $V_p$ in the $-x$ direction under an externally imposed temperature gradient, characterized by magnitude $G$ and oriented in the $-x$ direction. The FTA describes the temperature field directed along the interface growth direction ($+x$) as
\begin{equation}
T = T_0 + G\left(x - V_{\mathrm{iso}} t\right), \label{T_FTA}
\end{equation}
where $T_0$ is a reference temperature, chosen as the equilibrium liquidus temperature $T_M - m_e c_{\infty}$. The expression in Eq.~\eqref{T_FTA} is derived in a local frame of reference, hereafter referred to as the ``material frame'', which remains stationary with respect to the solidifying material. Within this frame, isotherms move at a constant speed $V_{\mathrm{iso}}$ in the $+x$ direction, where $V_{\mathrm{iso}}$ has the same magnitude as $V_p$. Tracking the solid-liquid interface is more convenient within the material frame, especially for a small simulation domain that encloses the interface. In a simulation using FTA, the temperature term in any of the PF equations is written as
\begin{equation}
\frac{T-T_M}{m_e}=\frac{G\left(x-x_0-V_{\mathrm{iso}} t\right)}{m_e}-c_{\infty},
\end{equation}
where $x_0$ is the location that initially coincides with the reference temperature $T_0$.

\paragraph*{Thermal Field Calculation}

During the solidification process, latent heat is released at the solid-liquid interface and diffuses away. To integrate TFC with the PF simulation, the temperature is modeled as a spatiotemporal scalar field $T(\mathbf{r}, t)$. Here, we consider the thermal diffusion in a frame of reference that is moving at $V_{\mathrm{iso}}$ along the $+x$ direction with respect to the material frame, which is hereafter referred to as the ``moving frame'' (also called gradient frame or lab frame in literature). Within this moving frame, the temperature field's evolution is governed by
\begin{equation}
\partial_{t} T = V_{\mathrm{iso}} \partial_{x} T + D_{T} \nabla^{2} T + \frac{\Delta h_{f}}{c_{p}} \frac{1}{\mathcal{V}} \int \frac{1}{2} \frac{\partial \phi}{\partial t} d\mathcal{V}, \label{dT_dt}
\end{equation}
where $D_T$ denotes the thermal diffusivity, $\Delta h_f$ represents the alloy's latent heat of fusion per unit volume, and $c_p$ is its heat capacity. The terms on the right-hand side of Eq.~\eqref{dT_dt} correspond, respectively, to heat convection by bulk motion, thermal diffusion (assumed uniform in both phases), and local heat generation from latent heat release, integrated over a small volume $\mathcal{V}$. While the PF and concentration equations are formulated in the material frame, the thermal equation is solved within an adiabatic zone in the moving frame. This necessitates the calculation of the thermal field in the moving frame and interpolating the local temperature back to the material sample frame. Given that $D_T \gg D_l$, solving the thermal equation \eqref{dT_dt} demands more stringent conditions for numerical stability. To tackle this numerical challenge, we develop a computationally efficient finite-difference algorithm that employs both a coarser grid and multiple time steps to solve Eq.~\eqref{dT_dt}. Precise synchronization of the boundary conditions is essential to link the coarser thermal grid with the finer grid for PF and concentration field. Details about the numerical implementation of TFC can be found in Appendix \ref{Appendix:TFC}.

\subsection{Remarks on model selection}

We have presented multiple quantitative PF models and their variations for rapid alloy solidification based on a similar variational framework. A discussion on the appropriate selection of a model for PF simulations is necessary.
For dilute alloys, both models I and II are applicable; however, there may be an unsafe region for Model I depending on the alloy and simulation parameters. For a specified $k_e$ value, the unsafe region of Model I corresponds to a large value of $\tilde{\lambda}$. As inferred from Eq.~\eqref{lambda}, this region is typically characterized by a high nominal concentration $c_{\infty}$, a small $k_e$, and/or a large $S$. While Model II addresses the fixed-point problem, it also establishes a higher energy barrier between the two bulk phases at $\phi = \pm 1$. As a result, the solute concentration profile across the diffuse interface becomes narrower, requiring finer grid spacing to resolve the interface accurately. This constraint becomes increasingly stringent for $S > 1$, making simulations with Model II significantly more time-consuming. Hence, we primarily utilize Model I for PF simulations of dilute alloys, resorting to Model II only when parameters fall into the unsafe region. Unless otherwise stated, all analyses and PF simulations for dilute alloys in this paper are performed based on Model I. 
For modeling concentrated alloys, only the generalized Model II presented in Sec.~\ref{Sec:ModelB_general} is used. The integration of complete free-energy functions for both solid and liquid phases allows for modeling arbitrary conditions within the two-phase region of a phase diagram. 
Additionally, the variation models, including the model without excess surface diffusion presented in Sec.~\ref{Sec:surface_diffusion} and the axisymmetric PF model in Appendix \ref{Appendix:Axisymm_PF}, can be based on either Model I or II, although our discussion focuses only on building these variation models based on Model I. The usage of the variation models is limited to specific scenarios that will be discussed subsequently. Moreover, the thermal condition TFC can be coupled to all the aforementioned PF models. Since TFC requires more simulation time than FTA, it is used only when the consideration of latent-heat diffusion becomes important.

\section{Phase-field solutions of solute trapping and asymptotic analyses} \label{Sec:Asymptotic_and_PF_solutions}

In this Section, we first derive the steady-state solutions for the models of dilute alloys in Sec.~\ref{Sec:PF_1D_dilute} and the model of concentrated alloys in Sec.~\ref{Sec:PF_1D_Al_Ag}. Then, we perform asymptotic analyses and show how the variational PF formulation matches the sharp-interface CG model assuming a hyperbolic tangent PF profile and a large $V$ limit.

\subsection{Steady-state solutions} \label{Sec:PF_solutions_1D}

\subsubsection{Dilute alloys} \label{Sec:PF_1D_dilute}

\paragraph{Solutions of Model I}

To obtain the steady-state solutions of PF and concentration profiles, we rewrite Eqs.\ \eqref{model_A_phi}-\eqref{model_A_c} in a frame moving with the interface at velocity $V$ in the $x$ direction:
\begin{align}
-\tau_0 V \frac{d \phi}{d x} &= (SW_0)^{2} \left( \frac{d^2 \phi}{dx^2} + \frac{1}{R_{i}} \partial_x \phi \right) + \phi - \phi^3 \nonumber \\
&- \lambda g^{\prime}(\phi)\left[c+\frac{(T-T_M)}{m_e} e^{b(1+g(\phi))}\right], \label{eq1} \\
- V \frac{dc}{dx} &= \frac{d}{dx} \left(D_l q(\phi) c \frac{d}{dx}[\ln c-b g(\phi)] \right). \label{eq2}
\end{align}
The term involving the radius of the interface curvature, $R_i$, comes from the Laplacian expansion in polar coordinates for a slightly curved moving interface. This Laplacian expansion supports the subsequent discussion on interface capillarity. Assuming this interface to be normal to the $x$ direction, our discussion can be confined to 1D along the $x$ axis.

By integrating both sides of Eq.~\eqref{eq2} with respect to $x$ and applying the boundary condition $c(\pm \infty)=c_\infty$ that ensures mass conservation, we obtain:
\begin{equation}
\frac{d c}{d x} = (c_\infty-c) \frac{V}{D_l q(\phi)} + b\, c\, \frac{d g(\phi)}{d x}. \label{c_scale}
\end{equation}
Furthermore, by multiplying both sides of Eq.~\eqref{eq1} by $d\phi/dx$ and integrating over $x$ from $-\infty$ to $+\infty$, we derive a self-consistent expression for the velocity-dependent temperature:
\begin{align}
T(V) = T_{M} &- \frac{b\, m_e}{1-k_e} \int_{-\infty}^{\infty} dx\,g^{\prime}(\phi) \,c\, \frac{d\phi}{dx} \label{T_PF} \\
&+ \frac{a_1 b \,m_e}{\lambda(1-k_e)} \frac{S W_0}{R_i} + \frac{a_1 b \,m_e}{\lambda(1-k_e)} \frac{\tau_0 V}{SW_0}, \nonumber
\end{align}
where we have defined
\begin{equation}
a_1 = SW_0\int_{-\infty}^{\infty} dx (d\phi/dx)^2. \label{a_1}
\end{equation}
The steady-state profiles $\phi(x)$ and $c(x)$ are uniquely determined by solving Eqs.\ \eqref{eq1} and \eqref{c_scale} with $T$ given by Eq.~\eqref{T_PF} and the boundary condition $c(\pm \infty) =c_{\infty}$. A ``full solution'' to this system is numerically obtained, which dynamically relaxes to a stationary state within the moving frame as detailed in Appendix \ref{Appendix:Full_PF_solution}. An ``approximate solution'' very close to the full solution can be obtained by assuming the PF profile for a moving interface remains close to its stationary profile $\phi_0(x)$, as defined in Eq.~\eqref{phi_0_x}. Under this approximation, the concentration profile is solely determined by Eq.~\eqref{c_scale}, which can be readily solved through numerical integration, yielding the concentration profiles shown in Fig. \ref{fig:q_phi}(b). Subsequently, $T(V)$ is determined using these $c(x)$ profiles and $\phi(x)=\phi_0(x)$. The corresponding functions $k(V)$ and $m(V)$ are obtained from the sharp-interface relations in Eq.~\eqref{Gibbs_Thomson} and the definition $k(V)=c_s/c_l$, where $c_s$ and $c_l$ are the concentrations on the solid and liquid sides of the interface, respectively, corresponding to $c_\infty$ and the peak value of $c(x)$. In order to match the velocity-dependent temperature relation from Eq.~\eqref{T_PF} in the PF model to the sharp-interface Gibbs-Thomson relation from Eq.~\eqref{Gibbs_Thomson}, we choose the model coefficients $\tau_0$ and $\lambda$ following Eqs.\ \eqref{tau0}-\eqref{lambda}. By matching the second and fourth terms on the right-hand side of Eqs.\ \eqref{Gibbs_Thomson} and \eqref{T_PF}, we obtain
\begin{equation}
\frac{m(V)}{m_e} = \frac{b}{(1-k_e)c_l} \int_{-\infty}^{\infty} dx g^{\prime}(\phi) c \frac{d\phi}{dx}, \label{m_v_m_e}
\end{equation}
and $\mu_k = \mu_k^0 a_1^0/a_1$, respectively. Matching the third term for the interface capillarity using the relation $\mathcal{K}=1/R_i$ for a 2D curved interface yields $\Gamma = \Gamma_0 a_1/a_1^0$. With the approximate solution, we get $\mu_k=\mu_k^0$ and $\Gamma = \Gamma_0$.

\begin{figure}[htbp!]
\centering
\includegraphics[scale=1]{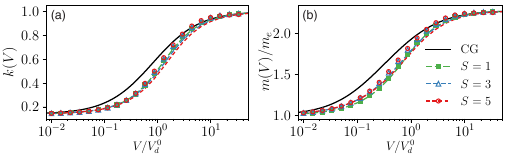}
\caption{
(a) $k(V)$ and (b) $m(V)$ functions derived from the full (symbols) and approximate (dashed lines) solutions for Model I. The black solid lines in (a) and (b) represent the CG model with coefficients determined in the large-velocity asymptotic limit.
\label{fig:Model_A_km}
}
\end{figure}

To obtain values of $A$ in Eq.~\eqref{q_phi} that yield $S$-independent trapping properties, we first calculate reference $k(V)$ and $m(V)$ curves with $S=1$ and $A=1$. For a given $S>1$, we then compute $k(V)$ and $m(V)$ curves for different $A$ values and select the $A$ value that minimizes the deviation from the reference curves over a large velocity range. This procedure, implemented using the approximate solution for Al-Cu alloy parameters, yields $A=6$ and 12 for $S=3$ and 5, respectively. The $k(V)$ and $m(V)$ plots obtained from the approximate and full solutions of the steady-state concentration and PF profiles for different $S$ and corresponding $A$ values are illustrated in Figs.\ \ref{fig:Model_A_km}(a) and (b). While the approximate solution is solely determined by $k_e$, the full solution, influenced by other alloy parameters, exhibits minor differences due to deviations of $\phi$ from $\phi_0$ at larger $V/V_d^0$. Notably, even with a single optimized parameter $A$ for each $S$, both $k(V)$ and $m(V)$ appear nearly $S$-independent over a large range of $V$. Despite the concentration profiles being $S$-dependent [Fig. \ref{fig:q_phi}(b)], they exhibit almost identical peak values that define $k(V)$. These profiles also yield nearly identical $m(V)$ values through Eq.~\eqref{m_v_m_e}.

\paragraph{Solutions of Model II}

\begin{figure}[htbp!]
\includegraphics[scale=1]{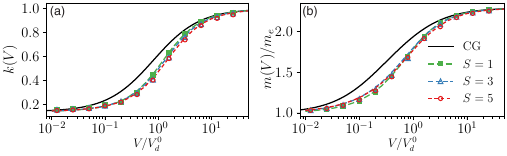}
\centering
\caption{
(a) $k(V)$ and (b) $m(V)$ functions obtained from the full (symbols) and approximate (dashed lines) solutions for Model II. The black solid lines in (a) and (b) represent the CG model with coefficients derived in the large-velocity asymptotic limit. Numerical calculations for the full solutions of Model II require smaller grid spacing, where $\Delta x/W = 0.6$ is used for $S = 1$, and $\Delta x/W = 0.4$ is used for $S = 3$ and $5$. \label{fig:Model_B_km}
}
\end{figure}

Given that the evolution equation for concentration in Model II is the same as that of Model I, Eq.~\eqref{c_scale} remains unchanged, and we only need to derive the solvability condition. We rewrite Eq.~\eqref{phi_newfT} in a frame moving with the interface at velocity $V$ in the $x$ direction:
\begin{align}
-\tau_0 V \frac{d \phi}{d x} &= (SW_0)^{2} \left( \frac{d^2 \phi}{dx^2} + \frac{1}{R_{i}} \partial_x \phi \right) + \phi - \phi^3  \label{phi_newfTx} \\
&- \lambda g^{\prime}(\phi)\left[c+\frac{(T-T_M)}{m_e} \frac{(k_e-1)}{2b} \right].    \nonumber
\end{align}
By integrating both sides of Eq.~\eqref{phi_newfTx}, we obtain the same solvability condition as in Eq.~\eqref{T_PF}. This suggests that, with the same choices of model coefficients following Eqs.\ \eqref{tau0}-\eqref{lambda}, Model II also matches to the sharp-interface Gibbs-Thomson relation through the relations $\mu_k = \mu_k^0 a_1^0/a_1$ and $\Gamma = \Gamma_0 a_1/a_1^0$. 
Furthermore, the same solvability condition suggests that the choices of $A$ in the quadratic form of $q(\phi)$ in Eq.~\eqref{q_phi} should be consistent between models I and II for a given alloy. Hence, we use the relation $k(V)=c_s/c_l$ and Eq.~\eqref{m_v_m_e} to compute the $k(V)$ and $m(V)$ curves for Al-Cu parameters, where the values $A=6$ for $S=3$ and $A=12$ for $S=5$ remain unchanged. Consequently, both $k(V)$ and $m(V)$ exhibit approximate $S$-independence across a broad range of $V$, as shown in Fig.~\ref{fig:Model_B_km}.

\paragraph{Phase-field profiles}

\begin{figure}[htbp!]
\includegraphics[scale=1]{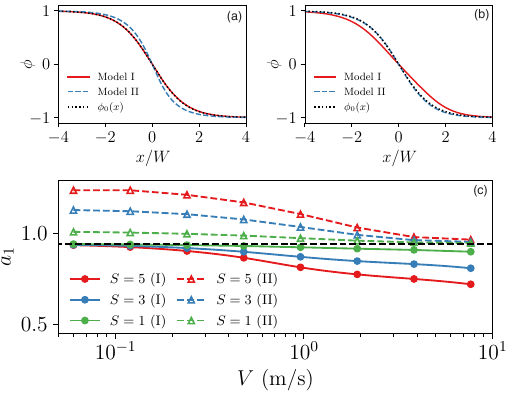}
\centering
\caption{
Comparison of 1D PF profiles from Model I and Model II with $S=5$ versus the hyperbolic tangent solution (represented by dashed lines) for velocities: (a) $V=0.06$ m/s and (b) $V=7.68$ m/s. (c) The computed $a_1$ values as a function of $V$ for both models with different interface thickness $S$. The black dashed line indicates $a_1^0=2 \sqrt{2} /3$, corresponding to the hyperbolic tangent profile $\phi_0(x)$. \label{fig:a_1}
}
\end{figure}

Both the kinetic coefficient $\mu_k = \mu_k^0 a_1^0 / a_1$ and the surface tension $\gamma = \gamma_0 a_1/a_1^0$ depend on the value of $a_1$ that is defined in Eq.~\eqref{a_1}. The latter relation is obtained since the Gibbs-Thomson coefficient is proportional to the surface tension, i.e., $\Gamma \sim \gamma$. Thus, the deviation of the PF profile from its hyperbolic tangent solution can influence both the interface kinetics and surface tension. Here we examine the PF profiles obtained from the full solutions of both Model I and Model II.
As shown in Fig.~\ref{fig:a_1}(a) for a relatively low velocity of $V=0.06$ m/s, the PF profile of Model I aligns closely with $\phi_0$, and the PF profile of Model II shows noticeable deviation. Conversely, at a higher velocity of $V=7.68$ m/s, as shown in Fig.~\ref{fig:a_1}(b), the PF profile of Model II closely matches $\phi_0$, while that of Model I deviates.
To gain more insights into these deviations, we plot $a_1$ defined in Eq.~\eqref{a_1} as a function of $V$ for both models with different values of $S$, as illustrated in Fig.~\ref{fig:a_1}(c). The results indicate that Model I's deviation amplifies with increasing $V$, whereas Model II's deviation is more pronounced at lower $V$. Additionally, for both models, the extent of deviation increases with $S$.


\subsubsection{Concentrated alloys} \label{Sec:PF_1D_Al_Ag}

We follow a procedure similar to that used for dilute alloys to obtain the steady-state solutions for concentrated alloys. By rewriting Eq.~\eqref{dcdt_B_general} in a frame moving with the interface at velocity $V$ in the $x$ direction, and integrating both sides once with respect to $x$ using the boundary condition $c(\pm \infty) = c_{\infty}$ imposed by mass conservation, we obtain:
\begin{align}
2 h_0 V & (c_{\infty}-c)=D_l q(\phi) c(1-c) \left[g^{\prime}(\phi)\left(\partial_c f_s-\partial_c f_l\right) \frac{\partial \phi}{\partial x} \right. \nonumber \\
&\left.+\left([1-g(\phi)] \partial_c^2 f_l+[1+g(\phi)] \partial_c^2 f_s\right) \frac{\partial c}{\partial x}\right]. \label{integral_c}    
\end{align}
Similarly, by rewriting Eq.~\eqref{dpdt_B_general} in the same moving frame and multiplying both sides by $d\phi/dx$ and integrating with respect to $x$, we obtain for a flat interface:
\begin{equation}
\tau_0 V \int(\partial_x \phi)^2 d x = \frac{1}{2h} \int g^{\prime}(\phi) \left[ f_s(c,T) - f_l(c,T) \right] \partial_x \phi d x. \label{integral_p}
\end{equation}
Unlike the case of dilute alloys, the first condition in Eq.~\eqref{integral_c} alone is insufficient to solve for the concentration profile $c_0(x)$ assuming a hyperbolic tangent PF profile, as both $T$ and $c_0(x)$ are undetermined. To obtain the approximate solution with $\phi(x) = \phi_0(x)$, we numerically solve a system consisting of both Eqs.~\eqref{integral_c} and \eqref{integral_p}. For a given $V$ and solidus concentration $c_{\infty}$, we first solve Eq.~\eqref{integral_c} to obtain $c_0(x)$ corresponding to various values of $T$ ranging from 800~K to the melting temperature $T_M = 933$~K. Then, by substituting these trial solutions into Eq.~\eqref{integral_p}, a unique temperature $T$ is found at the intersection point where the right-hand side of Eq.~\eqref{integral_p} equals its left-hand side. With the solutions of $T$ and $c_0(x)$ for given $V$ and $c_{\infty}$, the liquidus concentration at the same temperature is determined by the peak of the $c_0(x)$ profile. While we focus on the approximate solution of the generalized Model II here, its full solution can also be obtained through a similar procedure as in Appendix \ref{Appendix:Full_PF_solution}.

\begin{figure}[htbp!]
\includegraphics[scale=0.67]{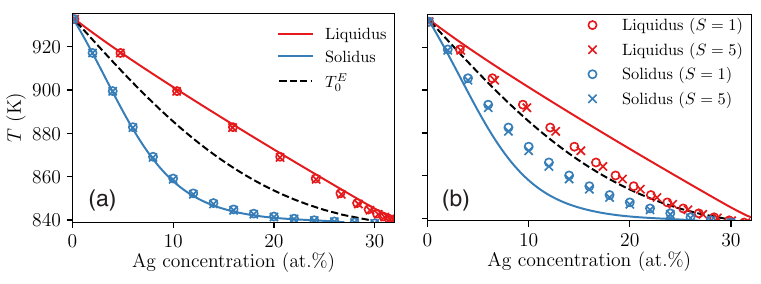}
\centering
\caption{
Equilibrium Al-Ag phase diagram obtained from CALPHAD (lines) and meta-stable phase diagrams obtained from 1D PF solutions with $S=1$ and $5$ (symbols) at finite interface velocities (a) $V$ = 0.03 m/s and (b) 1 m/s.
\label{fig:Nonlinear_Al_Ag}
}
\end{figure}

We investigate the Al-Ag alloy system with Gibbs free energy data for liquid and solid phases given in Eqs.~\eqref{fl_Al_Ag} and \eqref{fs_Al_Ag}, respectively. By scanning a range of Ag concentrations and solving for the corresponding liquidus concentrations and temperatures, we obtain nonequilibrium phase diagrams at various interface velocities $V$, as shown in Fig.~\ref{fig:Nonlinear_Al_Ag}. The phase diagrams derived from the 1D PF solutions with $S = 1$ and $5$ (symbols) are compared to the solidus and liquidus from CALPHAD (curves). As depicted in Fig.~\ref{fig:Nonlinear_Al_Ag}(a), the PF solutions at $V=0.03$ m/s closely match the equilibrium phase diagram derived from CALPHAD. Furthermore, as shown in Fig.~\ref{fig:Nonlinear_Al_Ag}(b) for $V = 1$ m/s, both the liquidus and solidus lines shift towards the $T_0^E$ line, defined as the line of equal free energy between the two phases at equilibrium. This shift is a manifestation of the solute trapping effects, which become significant at higher $V$. At this velocity, the kinetic undercooling shifts both the liquidus and solidus by 2 K with the $\mu_k^0$ given in Table \ref{tab:table1}, which is small compared with the temperature variation in the phase diagram. Additionally, the PF solutions with $S=1$ and $5$ agree well at lower $V$ and show only minor differences at higher $V$. This demonstrates that the strategy of enhancing solute diffusivity to upscale interface thickness also works effectively for the generalized Model II integrating complex free-energy functions. Here, we used $A=12$ in Eq.~\eqref{q_phi} for $S=5$ that is optimized for the Al-Cu alloy in the dilute limit. Remarkably, the same $A$ value continues to perform well for concentrated Al-Ag alloys without further optimization.

\subsection{Asymptotic analyses} \label{Sec:Asymptotic}

In this section, we expand the CG model equations \eqref{CGM_k}-\eqref{m_me} in the large-velocity asymptotic limit. Subsequently, we derive the CG model coefficients, including the diffusive velocity $V_d$ and the solute drag coefficient $\alpha$, and establish their connections with the coefficients in the PF model. The specific case under consideration is when $S=1$, with $W=W_0$ and $q(\phi)=(1-\phi)/2$. Since Eq.~\eqref{c_scale} is the same in both Model I and II, the following analyses apply to either model.

In the large-velocity limit, $V \gg V_d$, Eq.~\eqref{CGM_k} is expanded to
\begin{equation}
k(V) \approx 1 - (1 - k_{e})\left[\frac{V_{d}}{V}\right] + \mathcal{O}\left(\left[\frac{V_{d}}{V}\right]^{2}\right). \label{k_cg_approx}
\end{equation}
Similarly, Eq.~\eqref{m_me} is expanded to
\begin{equation}
\frac{m(V)}{m_e} \approx \frac{\ln(1/k_e)}{1-k_e} + (\alpha - 1) \ln\left(\frac{1}{k_e}\right) \left[\frac{V_{d}}{V}\right] + \mathcal{O}\left(\left[\frac{V_{d}}{V}\right]^{2}\right). \label{m_v_limit}
\end{equation}

We first derive a relation between the diffusive velocity $V_d$ in the CG model and the parameter $V_d^0 \equiv D_l/W_0$ defined within the PF model in the large-velocity asymptotic limit following the known method in Ref.~\cite{ahmad_solute_1998}. In the limit of $V \gg V_d^0$, we solve Eq.~\eqref{c_scale} and obtain
\begin{equation}
c(\phi) \approx c_{\infty}\left[1- \frac{V_{d}^0}{V} \frac{\ln1/k_e}{2} q(\phi) \frac{d g(\phi)}{d x}\right], \label{solution1}
\end{equation}
where $x$ is normalized by the interfacial thickness $W$. Given the 1D PF profile $\phi_0(x)$ from Eq.~\eqref{phi_0_x}, and its derivative, $\partial_x \phi_0 = -(1-\phi_0^2)/\sqrt{2}$, the concentration profile is solely determined by Eq.~\eqref{solution1}. Substituting $\phi(x) = \phi_0(x)$ into Eq.~\eqref{solution1} yields
\begin{equation}
c(\phi_0) \approx c_{\infty}\left[1+\frac{V_d^0}{V} \frac{15 \ln 1 / k_{e}}{32 \sqrt{2} }\left(1-\phi_{0}\right)^{4}\left(1+\phi_{0}\right)^{3} \right]. \label{c_phi0}
\end{equation}
The maximum value of $c(\phi_0)$ is found at $\phi_0 = -1/7$, Hence, the partition coefficient is derived from Eq.~\eqref{c_phi0} as $k=c_{\infty}/c_l=c_{\infty}/c(-1/7)$. By comparing this derived partition coefficient with Eq.~\eqref{k_cg_approx}, we obtain the relation between $V_d$ and $V_d^0$:
\begin{equation}
V_d=\frac{207360 \sqrt{2} \ln 1 / k_{e}}{823543(1-k_e)} V_d^0 \approx 0.356 \frac{ \ln 1 / k_{e}}{(1-k_e)} V_d^0. \label{V_d_limit}
\end{equation}
The prefactor is a function of $k_e$. For $k_e=0.14$, $V_d \approx 0.814 V_d^0$.

\begin{figure}[htbp!]
\includegraphics[scale=0.67]{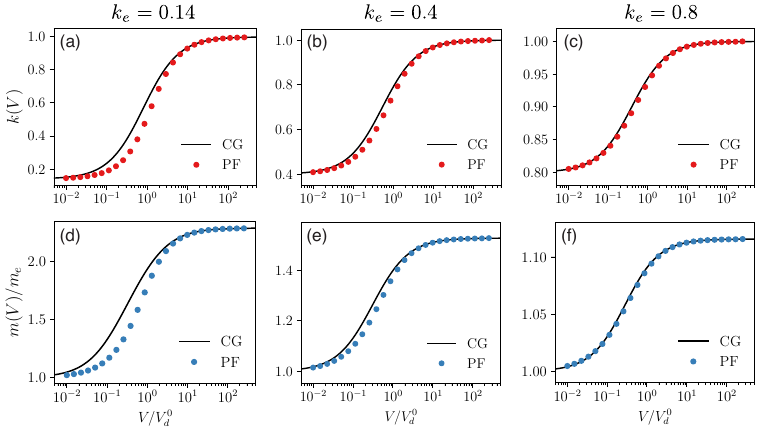}
\centering
\caption{
Comparison of partition coefficient $k(V)$ (a)-(c) and liquidus slope $m(V)$ (d)-(f) in the continuous growth (CG) model and the approximate solution of the PF model for $k_e=0.14$, 0.4, and 0.8.
\label{fig:km_ke}
}
\end{figure}

In Fig.~\ref{fig:km_ke}(a)-(c), we compare $k(V)$ in the CG model and the approximate solution of the PF model assuming $\phi(x)=\phi_0(x)$ for different $k_e$ values. For $k(V)$ curves of the CG model, we compute $V_d$ using Eq.~\eqref{V_d_limit} with $k_e$ as an input parameter. The PF model's approximate solutions are obtained through the numerical calculation of Eq.~\eqref{c_scale}. For $k_e = 0.14$, the PF model's approximate solution agrees with the CG model predictions for both low ($V \ll V_d^0$) and high ($V \gg V_d^0$) velocity regimes, albeit discrepancies emerge at intermediate velocities. Notably, as $k_e$ increases from 0.14 to 0.8, the models exhibit almost perfect agreement over the entire spectrum of velocities.

Meanwhile, we use Eq.~\eqref{m_v_m_e} to find the nonequilibrium liquidus slope in the large-velocity asymptotic limit, where both PF and concentration profiles are required to evaluate the integral. As before, we assume the PF profile $\phi(x)=\phi_0(x)$. For the concentration profile, we substitute Eq.~\eqref{V_d_limit} into Eq.~\eqref{c_phi0} and obtain
\begin{equation}
c(\phi_0) \approx c_{\infty}\left[1+\frac{V_d}{V} \frac{15 (1-k_e)}{32 \sqrt{2} C_1 }\left(1-\phi_{0}\right)^{4}\left(1+\phi_{0}\right)^{3}\right], \label{c_phi_C1}
\end{equation}
where $C_1={207360 \sqrt{2}}/{823543}$. According to Eq.~\eqref{m_v_m_e} and the relationship $k = c_{\infty}/c_{l} \approx 1 - (1 - k_{e})V_{d}/V$, we derive in the $V \gg V_d$ limit that
\begin{align}
\frac{m(V)}{m_e} \approx & \frac{\ln 1/k_e}{1-k_e}+\left(\frac{25}{77 \sqrt{2}C_1}-1 \right) \ln 1/k_e \left[\frac{V_{d}}{V}\right] \label{m_v_m_e2} \\
&+\mathcal{O} \left(\left[\frac{V_{d}}{V}\right]^{2}\right). \nonumber
\end{align}
By comparing Eq.~\eqref{m_v_m_e2} with Eq.~\eqref{m_v_limit}, we readily identify the solute drag coefficient $\alpha = 25/(77\sqrt{2}C_1) \approx 0.645$, which is close to the asymptotic value $\alpha \approx 0.686$ reported in Ref.~\cite{ahmad_solute_1998}.
In the asymptotic analysis of the current PF model, as indicated by Eqs.~\eqref{solution1}-\eqref{c_phi_C1}, the value of $\alpha$ is independent of $k_e$ and is solely determined by the interpolation functions $q(\phi)$ and $g(\phi)$. Thus, these interpolation functions could potentially be further tuned to model different amounts of solute drag for specific alloy systems where this effect has been quantified experimentally or by atomistic simulations \cite{yang2011atomistic,kavousi2020interface,antillon2023solute}.

In Fig.~\ref{fig:km_ke}(d)-(f), we compare $m(V)$ in the CG model and the approximate solution of the PF model for different $k_e$. For the $m(V)$ curves of the CG model, we calculate $V_d$ via Eq.~\eqref{V_d_limit}, while $\alpha$ is held constant at 0.645, as indicated by the asymptotic analyses. The PF model's $m(V)$ curves are obtained using Eq.~\eqref{m_v_m_e} with $\phi(x)=\phi_0(x)$ and the approximate solution of the concentration field [solution of Eq.~\eqref{c_scale} with $\phi(x)=\phi_0(x)$] as inputs. Similar to the case of $k(V)$, an almost perfect $m(V)$ agreement is found over the entire velocity range as $k_e$ increases from 0.14 to 0.8. 

Given that the PF model accounts for a spatially diffuse interface region and the CG model employs a sharp-interface description, the quantitative agreement between the two models across the entire velocity range for varying $k_e$ values is not generally expected. Nevertheless, the agreement between them approaches near perfection as $k_e$ values increase. Within the scope of our analysis, the PF predictions of $k(V)$ and $m(V)$ for a realistic solid-liquid interface width ($S=1$) are posited as the proximate ``ground truth''.

\section{Simulation results} \label{Sec:Numerical_results}

\begin{table*}[htbp!]
\caption{\label{tab:table1}
Materials, process, and simulation parameters.
}
\begin{ruledtabular}
\begin{tabular}{lp{5.7cm}llll}
Symbol & Description & Al-Cu & Al-Ag & Unit & \\
\colrule
$D_l$ & Solute diffusivity in liquid & 2400 & 3000 & $\mathrm{\mu m^2 \, s^{-1}}$ & \cite{lee_diffusion-coefficient_2004} \\
$k_e$ & Equilibrium partition coefficient & 0.14 & – & & \cite{kurz1989fundamentals} \\
$m_e$ & Equilibrium liquidus slope & $2.6$ & – & K wt.\%$^{-1}$ & \cite{clarke_microstructure_2017} \\
$\Gamma_0$ & Gibbs-Thomson coefficient & 0.24 & 0.196 & $\mathrm{\mu m \, K}$  & \cite{kurz1989fundamentals,gunduz_measurement_1985} \\
$W_0$ & Atomic interface thickness & 1 & 1 & nm\\
$V_d^0$ & Diffusive speed $D_l/W_0$ & 2.4 & 3.0 & $\mathrm{m \, s^{-1}}$ \\
$\mu_k^0$ & Interface kinetic coefficient & 0.5 & 0.5 & $\mathrm{m \, s^{-1} \, K^{-1}}$ & \cite{mendelev_molecular-dynamics_2010} \\
$\epsilon_k$ & Kinetic anisotropy strength & 0.1 & 0.1 \\
$\epsilon_s$ & Interface free-energy anisotropy strength & 0.012 & 0.012 \\
\\
$G$ & Temperature gradient & $5 \times 10^6$ & – & K m$^{-1}$ & \cite{carrard_about_1992} \\
$D_T$ & Thermal diffusivity & $5.35 \times 10^{-5}$ & $5.35 \times 10^{-5}$ & $\mathrm{m^2 \, s^{-1}}$ & \cite{karma_interface_1993}\\
$T_M$ & Melting temperature of pure substance & 933 & 933 & K & \cite{karma_interface_1993} \\
$L/c_p$ & Latent heat of fusion per unit volume divided by the heat capacity & 340.5 & 340.5 & K & \footnote{Average value of solid and liquid phases for pure Al.}\\
\\
$W$ & Interface thickness & 1, 3, 5 & 1, 3, 5 & $W_0$\\
$\Delta x$ & Grid spacing & 0.8 & 0.6 & $W$ \\
$\Delta x_T$ & Coarser grid spacing for the thermal field & 10 & 17 & $\Delta x$ \\
\end{tabular}
\end{ruledtabular}
\end{table*}

In this section, we first present the evolution equations of the PF model with interfacial anisotropies and their numerical implementation in Sec.~\ref{Sec:Evolution_equations}. Then, we present 2D simulation results in Sec.~\ref{Sec:PF_simulation_2D} and 3D simulation results in Sec.~\ref{Sec:PF_simulation_3D}.

\subsection{Evolution equations and numerical implementation} \label{Sec:Evolution_equations}

In 2D and 3D PF simulations, we consider the anisotropies in both the excess free-energy of the solid-liquid interface and the interface kinetics. The choices of interface width $W(\mathbf{n})=W a_s(\mathbf{n})$ and time constant $\tau(\mathbf{n})=\tau_0 a_s(\mathbf{n})^2 / a_k(\mathbf{n})$ model general anisotropic forms of the excess free-energy of the solid-liquid interface $\gamma(\mathbf{n})=\gamma_0 a_s(\mathbf{n})$ and interface kinetic coefficient $\mu_k(\mathbf{n})=\mu_k^0 a_k(\mathbf{n})$, where $\mathbf{n}$ is the direction normal to the interface. In 2D, we consider a simple four-fold symmetry for the anisotropy functions 
\begin{align}
a_s (\theta)=1+\epsilon_s \cos (4 \theta), \\
a_k (\theta)=1+\epsilon_k \cos (4 \theta),
\end{align}
where $\theta=\tan ^{-1}(\partial_y \phi/\partial_x \phi)$ is the angle between $\mathbf{n}$ and the $x$ axis. The parameters $\epsilon_s$ and $\epsilon_k$ represent the anisotropy strengths for the interface free-energy and interface kinetics, respectively. In 3D, the anisotropy functions are
\begin{align}
a_s(\mathbf{n})&=\left(1-3 \epsilon_s\right)\left[1+\frac{4 \varepsilon_s}{1-3 \varepsilon_s}\left(n_x^4+n_y^4+n_z^4\right)\right], \label{as_3D} \\
a_k(\mathbf{n})&=\left(1-3 \epsilon_k\right)\left[1+\frac{4 \varepsilon_k}{1-3 \varepsilon_k}\left(n_x^4+n_y^4+n_z^4\right)\right], \label{ak_3D}
\end{align}
where $\mathbf{n}=\left(n_x, n_y, n_z\right)$ is the unit vector normal to the interface pointing towards the liquid, and $n_i=-\partial_i \phi / |\vec{\nabla} \phi|$.

To facilitate numerical implementation, we scale the length by $W$ and time by $\tau_0$. The final evolution equations incorporating both excess interface free-energy and kinetic anisotropies are given by
\begin{align}
\frac{a^2_s(\mathbf{n})}{a_k(\mathbf{n})} \frac{\partial \phi}{\partial t} &= \vec{\nabla} \cdot\left[a_s(\mathbf{n})^{2} \vec{\nabla} \phi\right] + \phi - \phi^{3} \label{Diemensionless_phi} \\
&+ \sum_{i}\left[\partial_{i}\left(|\vec{\nabla} \phi|^{2} a_s(\mathbf{n}) \frac{\partial a_s(\mathbf{n})}{\partial\left(\partial_{i} \phi\right)}\right)\right] \nonumber \\
&- \tilde{\lambda} g^{\prime}(\phi) \left[\tilde{c} + \widetilde{T} e^{b(1+g(\phi))} \right], \nonumber \\
\frac{\partial \tilde{c}}{\partial t} &= \widetilde{D}_l \vec{\nabla} \cdot \left \{ q(\phi) \tilde{c} \vec{\nabla}[\ln \tilde{c}-b g(\phi)] \right \}, \label{Diemensionless_c}
\end{align}
where $\tilde{\lambda} \equiv \lambda c_{\infty}$, $\tilde{c} \equiv c/c_{\infty}$, and $\widetilde{D}_l \equiv D_l \tau_0 / W^2 = D_l/ (\Gamma_0 \mu_k^0)$ are the dimensionless diffusion coefficient. The index $i$ runs over the spatial coordinates, i.e., $i=x,\,y$ in 2D and $i=x,\,y,\,z$ in 3D. For isothermal solidification, $\widetilde{T}=(T-T_M)/(m_e c_{\infty})$ is a constant. For directional solidification with FTA, we have
\begin{equation}
\widetilde{T} = \frac{x-x_0-\widetilde{V}_{\mathrm{iso}} t}{\widetilde{l}_T}-1,
\end{equation}
where $\widetilde{V}_{\mathrm{iso}} \equiv \tau_0 V_{\mathrm{iso}} /W $ and $\widetilde{l}_T \equiv l_T/W = (m_e c_{\infty} /G)/W$ denote the dimensionless pulling speed and thermal length, respectively. The scaled reference location $x_0$ coincides with the reference temperature $T_L - m_e c_{\infty}$. 

With the thermal condition given by TFC, Eq.~\eqref{dT_dt} is coupled with Eqs.\ \eqref{Diemensionless_phi}-\eqref{Diemensionless_c}, and the evolution equation of the $\widetilde{T}$ field has a dimensionless form
\begin{equation}
\partial_{t} \widetilde{T} = \widetilde{V}_{\mathrm{iso}} \partial_{x} \widetilde{T} + \widetilde{D}_{T} \nabla^2 \widetilde{T} + \frac{H}{\mathcal{V}} \int \frac{1}{2} \frac{\partial \phi}{\partial t} d\mathcal{V}, \label{heat_diff}
\end{equation}
where $\widetilde{D}_T \equiv D_T \tau_0 / W^2$ and $H \equiv \Delta h_{f} / (c_p m_e c_{\infty})$. While the numerical calculations for $\phi$ and $\tilde{c}$ fields are confined to a relatively small domain (3-6 $\mu$m in the $x$ direction) enclosing the solid-liquid interface, the thermal field calculation is performed within a much larger adiabatic zone (50 $\mu$m in the $x$ direction). To achieve an efficient computation of coupled evolution equations, we develop a coarser-grid and multi-step strategy for the finite-difference implementation of Eq.~\eqref{heat_diff}, with details provided in Appendix \ref{Appendix:TFC}.

PF simulations are carried out for directional solidification of dilute Al-Cu alloys and concentrated Al-Ag alloys. Unless otherwise specified, we choose $S=5$, and the other parameters are listed in Table~\ref{tab:table1}. The PF model is implemented on massively parallel graphic processing units (GPUs) utilizing the computer unified device architecture (CUDA) programming language. The dimensionless model equations \eqref{Diemensionless_phi}-\eqref{Diemensionless_c} are solved on a square lattice in 2D and a cubic lattice in 3D through a finite difference method for spatial derivatives and an Euler explicit time-stepping scheme. For leading differential terms including the Laplacian and divergence, we adopt isotropic discretizations as introduced in Ref.~\cite{ji_isotropic_2022}. For anisotropy terms in Eq.~\eqref{Diemensionless_phi}, we first analytically expand these terms to first and second derivatives of $\phi$, then solve them on a regular stencil following the procedure given in Ref.~\cite{tourret2015growth}. The explicit time step is set as $\Delta t/\tau_0 = R_t (\Delta x/W)^2$, with the coefficient $R_t$ being 0.15 and 0.1 in 2D and 3D simulations, respectively. These choices of $\Delta t$ ensure the numerical stability of the evolution equations for both the PF and concentration field when $D_l \tau_0/W^2 \le 1$. For $D_l \tau_0/W^2 > 1$, the stability condition is determined by the PF equation \eqref{Diemensionless_phi}. With TFC, the finite-difference implementation of Eq.~\eqref{heat_diff} employs a coarser grid and a finer time stepping (Appendix \ref{Appendix:TFC}).

In most simulations, the solid-liquid interface starts as a planar structure situated at the liquidus temperature in a stationary state. An initial perturbation $\eta \beta(\vec{r})$ is applied perpendicular to the interface location (along the $x$-axis), where $\eta=0.5\Delta x$ denotes the noise amplitude and $\beta(\vec{r})$, a function dependent on $\vec{r}$ along the horizontal interface location, is generated randomly from a uniform distribution between $[-0.5,\,0.5]$. The initial PF profile adopts the 1D approximate solution $\phi_0(x)$ from Eq.~\eqref{phi_0_x}, and the initial concentration profile is the corresponding equilibrium profile $c_0(x)$ at the liquidus temperature as given by Eq.~\eqref{c_0_x}. In the other cases, we also import the scalar fields from a previous simulation as the initial conditions.

\subsection{Phase-field simulations in two dimensions} \label{Sec:PF_simulation_2D}

\subsubsection{Dendrite shapes} \label{Sec:2D_dendrite_shapes}

We begin by examining the velocity range below the onset of banding, where stable dendritic array structures are formed. In 2D PF simulations using FTA, we obtain a single steady-state dendrite (SSD) for an Al-3wt.\% Cu alloy with $V_{\mathrm{iso}} = 0.12$ m/s and $G = 5 \times 10^6$ K/m. Within the simulation domain, periodic boundary conditions are applied in the $\pm y$ directions, and the width of the simulation domain along $y$ corresponds to the primary dendrite array spacing $\Lambda$. The value of $\Lambda=0.65$ $\mu$m we consider here is situated within the stable range of $\Lambda$. The lower boundary of this spacing stability is associated with an elimination instability, while its upper boundary is linked to tertiary branching or tip-splitting instabilities \cite{echebarria_onset_2010,song_thermal-field_2018}. At the $+x$ boundary, we set $\phi=-1$ and $\tilde{c}=1$. The most advanced solid-liquid interface (dendrite tip) is maintained at a fixed $x$ location by pulling back the entire simulation domain and truncating the excess portion of the PF and concentration field at the rear at the $-x$ boundary. Here, a no-flux boundary condition is implemented in the $-x$ direction.

\begin{figure}[htbp!]
\includegraphics[scale=1]{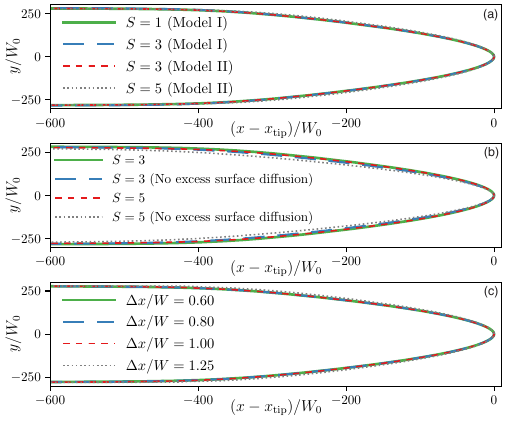}
\centering
\caption{
Comparison of dendrite morphologies from: (a) simulations utilizing both Model I and Model II with different interface thickness $S$; (b) simulations of Model I, both with and without surface diffusion for $S=3$ and 5; and (c) simulations of Model I with $S=5$ and different grid spacing $\Delta x$. All simulations were performed for the solidification of an Al-3wt.\% Cu alloy under conditions $G = 5 \times 10^6$ K/m and $V_{\mathrm{iso}} = 0.12$ m/s.
\label{fig:dendrite_shapes}
}
\end{figure}

In Fig.~\ref{fig:dendrite_shapes}(a), we compare the dendrite shapes from simulations using Model I and Model II. Simulations employing both models with different $S$ values produce nearly identical contours defined by $\phi=0$. For Model I, we utilize a grid spacing of $\Delta x/W=0.8$ in the finite-difference implementation. However, Model II requires a finer grid spacing of at least $\Delta x/W=0.6$ to prevent the dendritic growth becomes unstable due to the numerical reason. For the scenario with $S=3$, this means that the simulation using Model II demands at least double the computation time compared to Model I, even when using similarly optimized codes. Consequently, Model I is the favored choice when parameters reside within the safe zone as identified in Fig.~\ref{fig:fixed_point}.

Furthermore, surface diffusion and interface stretching are known to influence the pattern selection \cite{Karma2001,echebarria_quantitative_2004}. We employ the PF model formulated in Sec.~\ref{Sec:surface_diffusion} to eliminate the excess surface diffusion for $S>1$. As shown in Fig.~\ref{fig:dendrite_shapes}(b), a comparison between standard simulations using Model I and the PF model which eliminates excess surface diffusion for $S>1$ reveals minor differences for $S=3$. However, these differences become more pronounced for $S=5$, given that the amplified surface diffusion is more significant with a larger $S$. With the same $\Lambda$, simulations accounting for excess surface diffusion exhibit a wider dendrite shape, which is a consequence of the additional solute redistribution at the solid-liquid interface beneath below the dendrite tip. Although the simulations of Model I and the variation model in Sec.~\ref{Sec:surface_diffusion} exhibit relatively minor differences in dendrite shapes at lower velocities, their discrepancies become pronounced at higher velocities, where the latter model yields poor predictions on the stability of cellular/dendritic solutions. Since the spurious effects for $S>1$ are attributed to both excess surface diffusion and interface stretching, a more comprehensive exploration of their individual and combined effects is merited in future studies. 
While interface stretching can also be eliminated within the thin-interface limit for quasi-equilibrium growth conditions \cite{echebarria_quantitative_2004}, eliminating both surface diffusion and interface stretching across the entire $V$ range is considerably more challenging, which warrants further investigation.


Lastly, we compare dendrite shapes from simulations of Model I using varied grid spacings. As illustrated in Fig.~\ref{fig:dendrite_shapes}(c) for simulations with $S=5$, the dendrite shape convergence is excellent for $\Delta x/W < 1$, but the simulation with $\Delta x/W = 1.25$ begins to diverge. Unless otherwise specified, we have consistently adopted a conservative value of $\Delta x/W = 0.8$ for all Model I simulations throughout this paper.

\subsubsection{Microstructure development}

\begin{figure}[htbp!]
\includegraphics[scale=1]{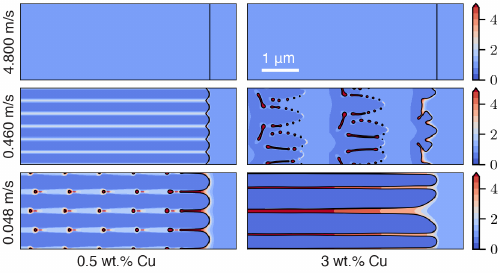}
\centering
\caption{
With frozen temperature approximation (FTA), microsegregation patterns ($\tilde{c}=c/c_\infty$ color maps) simulated for Al-0.5wt.\% Cu and Al-3wt.\% Cu alloys with $G=5\times 10^6$ K/m at different $V_{\mathrm{iso}}$. Black curves represent the solid-liquid interfaces.
\label{fig:FTA_2D}
}
\end{figure}

\begin{figure}[htbp!]
\includegraphics[scale=1]{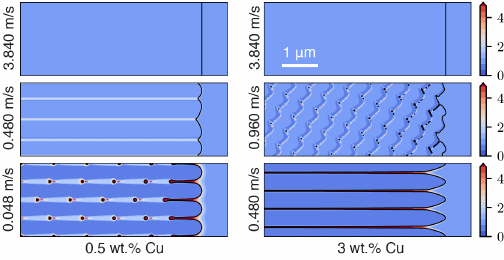}
\centering
\caption{
With thermal field calculation (TFC), microsegregation patterns ($\tilde{c}=c/c_\infty$ color maps) simulated for Al-0.5wt.\% Cu and Al-3wt.\% Cu alloys with $G=5\times 10^6$ K/m at different $V_{\mathrm{iso}}$. Black curves represent the solid-liquid interfaces.
\label{fig:TFC_2D}
}
\end{figure}

A series of 2D PF simulations were performed from initially planar interfaces to explore microstructure development under varying growth and thermal conditions. For binary alloys at very dilute concentrations, a transition from cellular-dendritic to planar front growth is typically observed \cite{ludwig_direct_1996,boettinger_simulation_1999}. This phenomenon is reproduced in our simulations of an Al-0.5wt.\% Cu alloy using both FTA and TFC, as illustrated in Figs.~\ref{fig:FTA_2D} and \ref{fig:TFC_2D}. At an isotherm velocity $V_{\mathrm{iso}} = 0.048$ m/s, dendritic array structures with shallow liquid grooves are formed under both thermal conditions, with heightened solute concentration in the grooves. As $V_{\mathrm{iso}}$ increases, these grooves gradually diminish, eventually transitioning to a planar interface at a critical isotherm velocity $V_c \approx 0.5$ m/s. In the case of a more concentrated Al-3wt.\% Cu alloy, simulations reveal dendritic arrays with deep liquid grooves below a critical isotherm velocity $V_{\mathrm{iso}}<V_{c,1}$. For $V_{\mathrm{iso}}>V_{c,1}$, an initially planar interface evolves into an oscillatory cycle, giving rise to banded microstructures. This banding phenomenon persists across a broad $V_{\mathrm{iso}}$ range, followed by restabilization to a planar morphology at much larger $V_{\mathrm{iso}}$. Since the onset of banding depends on the initial condition, another critical isotherm velocity, $V_{c,2}$, is found by progressively increasing $V_{\mathrm{iso}}$ from an SSD solution until banding emerges. For the Al-3wt.\% Cu alloy, $V_{c,1} \approx 0.45$ m/s and $V_{c,2} \approx 0.88$ m/s. Within the velocity range $V_{c,1} < V_{\mathrm{iso}} < V_{c,2}$, the interface dynamics is fundamentally bistable, i.e., the solution can be either dendritic or banded depending on initial conditions. The banded microstructures are also significantly influenced by the thermal condition. The FTA simulation (Fig.~\ref{fig:FTA_2D}) exhibits dendritic dark bands and microsegregation-free light bands growing parallel to the thermal axis. In contrast, the TFC simulation (Fig.~\ref{fig:TFC_2D}) shows bands growing at a slight angle relative to the thermal axis, attributable to the slowdown of lateral interface spreading due to latent-heat rejection \cite{ji_microstructural_2023}, resulting in significantly reduced band spacing from approximately 2 $\mu$m to 500 nm.

\begin{figure}[htbp!]
\includegraphics[scale=1]{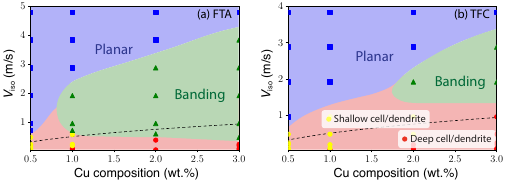}
\centering
\caption{
Al-Cu microstructure selection maps by 2D PF simulations with (a) frozen temperature approximation (FTA) and (b) thermal field calculation (TFC). The temperature gradient in all simulations is $G = 5 \times 10^6$ K/m. Dashed lines represent the analytical absolute stability limit $V_a$. The boundaries between different microstructural regions are crudely approximated based on sparse simulation results indicated by the symbols.
\label{fig:selection_map}
}
\end{figure}

Microstructure selection maps by 2D PF simulations from initially planar interfaces are presented in Fig.~\ref{fig:selection_map} for dilute Al-Cu alloys. Both selection maps under FTA and TFC conditions reveal a smooth transition from cellular-dendritic to planar front growth for dilute concentrations, and the banded microstructures are observed only for concentrations exceeding a critical value $c^*$. Linear stability analysis for the Al-Cu alloy indicates that $c^*=0.18$ wt.\% when using FTA and $c^*=2.07$ wt.\% when considering the latent-heat diffusion \cite{karma_interface_1993}. According to Fig.~\ref{fig:selection_map}, the critical concentrations for banding are approximately $c^* \approx 0.8$ wt.\% with FTA—a value larger than the analytically predicted one—and $c^* \approx 1.8$ wt.\% with TFC, which is slightly smaller than the predicted value. While directly comparing 2D PF simulations to 1D linear stability analysis might seem somewhat oversimplified, our results qualitatively reproduce key findings from the stability analysis. They demonstrate the transition from cellular-dendritic to planar front growth in dilute alloys, the emergence of banding within a specific velocity range for concentrations above $c^*$, and a larger $c^*$ when incorporating latent-heat diffusion.

\begin{figure}[htbp!]
\includegraphics[scale=0.65]{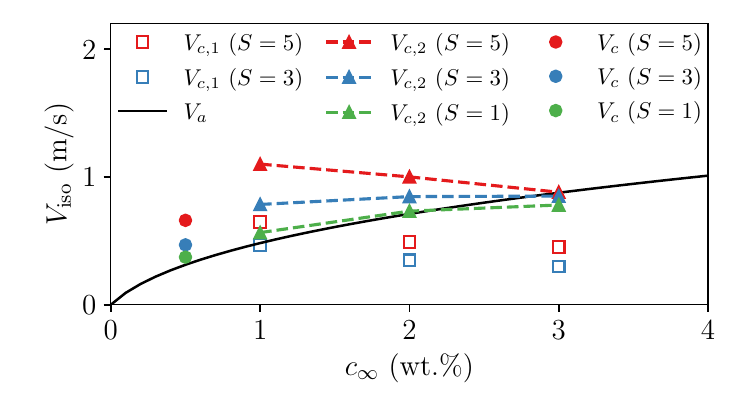}
\centering
\caption{
Critical isotherm velocities obtained from 2D PF simulations with different values of $S$ for microstructural transitions: from cellular/dendritic to planar ($V_c$), and from cellular/dendritic to banding ($V_{c,1}$ and $V_{c,2}$). Open symbols represent simulations initiated from planar interfaces with constant $V_{\mathrm{iso}}$, while filled symbols represent simulations initiated from steady-state solutions with progressively increasing $V_{\mathrm{iso}}$. The black line denotes the analytical absolute stability limit $V_a$.
\label{fig:Va_convergence}
}
\end{figure}

In Sec.~\ref{Sec:2D_dendrite_shapes}, we examined the convergence of dendritic shapes at low velocities in PF simulations with different values of $S$. Given that the transition to banded microstructures occurs at higher velocities, we also assess the convergence of PF simulations at these velocities. Fig.~\ref{fig:Va_convergence} shows the critical isotherm velocities for microstructural transitions obtained from 2D PF simulations with different $S$. The convergence of $V_{c,1}$—related to the transitions from cellular/dendritic to banding for simulations initiated from planar interfaces—is reasonably good across different alloy compositions. The convergence of $V_{c,2}$—related to the transitions from cellular/dendritic to banding for simulations starting from steady-state solutions with progressively increased $V_{\mathrm{iso}}$—is almost perfect for 3wt.\% Cu but becomes less consistent for lower alloy compositions, which is likely more influenced by excess surface diffusion and interface stretching at lower compositions. As alloy compositions become more dilute, simulations with $S=3$ show a narrowing velocity interval for banding until $V_{c,1} \approx V_{c,2} = V_c$ at an alloy composition between 0.5wt.\% and 1wt.\% Cu. Specifically, for Al-0.5wt.\% Cu, shallow cellular structures directly evolve into planar interfaces at $V_c$. The convergence of $V_c$ at 0.5wt.\% Cu is better than that of $V_{c,2}$ at 1wt.\% Cu, likely because oscillatory interface behavior during banding amplifies the spurious effects for $S>1$. Additionally, we compare the critical isotherm velocities from PF simulations to the absolute stability limit $V_a$, an analytical prediction of when the dendritic solution loses stability \cite{mullins_stability_1964,trivedi_morphological_1986,ludwig_direct_1996,boettinger_simulation_1999}, as implicitly defined by the equation
\begin{equation}
V_a = \frac{D_l m(V_a) c_{\infty}[1 - k(V_a)]}{k(V_a)^2 \Gamma}. \label{abs_stab}
\end{equation}
We used Eq.~\eqref{abs_stab} and the $k(V)$ and $m(V)$ curves computed from the approximate PF solution with $S=1$ to solve for $V_a$ in Fig.~\ref{fig:Va_convergence}. Notably, the upper velocity limits of cellular/dendritic solutions in PF simulations, denoted by $V_c$ and $V_{c,2}$ for different alloy compositions, agree well with $V_a$. Given that the PF model with $S=1$ does not introduce spurious effects associated with an unphysical interface thickness, simulation results with $S=1$ provide valuable references for understanding interface dynamics at high velocities.

\subsubsection{Simulations of concentrated alloys} \label{Sec:PF_2D_Al_Ag}

\begin{figure}[htbp!]
\includegraphics[scale=1]{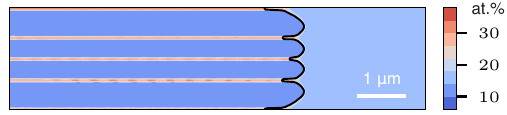}
\centering
\caption{
2D PF simulation ($S=3$) with thermal field calculation for Al-15at.\% Ag at $V$ = 0.075 m/s. 
Colormaps represent the Ag concentration in at.\%, and black curves represent the solid-liquid interfaces. \label{fig:Al_Ag_2D}
}
\end{figure}

The PF model incorporating the complete free-energy functions allows for simulating the rapid solidification of concentrated alloys under far-from-equilibrium conditions. As demonstrated in Fig.~\ref{fig:Nonlinear_Al_Ag}(a), the 1D PF solution with $S=1$ and 5 at a small $V$ reproduces well the equilibrium phase diagram from CALPHAD, where the Ag concentration spans from dilute to hypo-eutectic. Here, we carry out a 2D PF simulation with $S=5$ for the rapid solidification of Al-Ag alloys. The alloy and modeling parameters for Al-Ag are given in Table \ref{tab:table1}, and the free-energy database from Ref.~\cite{witusiewicz_agcu_2004} is summarized in Eqs.~\eqref{fl_Al_Ag}-\eqref{fs_Al_Ag}. We solve Eq.~\eqref{dcdt_B_general} and the anisotropic form of Eq.~\eqref{dpdt_B_general}, using the same excess interface free-energy and kinetic anisotropies given in Eq.~\eqref{Diemensionless_phi}. The thermal condition is TFC with details given in Appendix \ref{Appendix:TFC}.
We employed the same boundary conditions as outlined in Sec.~\ref{Sec:2D_dendrite_shapes}. The stationary solid-liquid interface is initially positioned at the liquidus temperature within an imposed temperature gradient. Alternatively, the planar PF solution at a finite $V$ and a specific $c_\infty$, obtained through a procedure in Sec.~\ref{Sec:PF_1D_Al_Ag}, yields a steady-state concentration profile and interface temperature, which can also be used as the initial condition with a small perturbation to the planar interface. This initial condition can be applied to enhance the stability of the simulation when the interface temperature is close to the eutectic point. Here, we use the first initial condition, and an illustrative example of a PF simulation for Al-15at.\% Ag at $V_{\mathrm{iso}}=0.075$ m/s is shown in Fig.~\ref{fig:Al_Ag_2D}. The simulation gives rise to a steady-state cellular array with shallow liquid grooves. These grooves form parallel channels in the microsegregation pattern, with solute concentrations reaching around 30at.\% Ag. Simulations extending beyond the absolute stability limit can also be conducted. A more comprehensive study comparing the PF modeling of Al-Ag alloys with experimental results will be presented in a separate study.

\subsection{Phase-field simulations in three dimensions} \label{Sec:PF_simulation_3D}

\subsubsection{Axisymmetric simulations}

\begin{figure}[htbp!]
\includegraphics[scale=1]{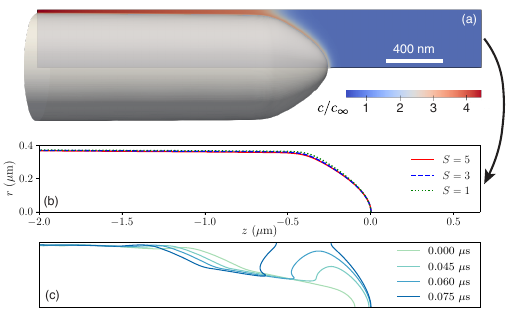}
\centering
\caption{
(a) Illustration of an axisymmetric simulation ($c/c_{\infty}$ colormap) alongside the 3D-reconstructed dendrite shape. (b) Comparison of steady-state dendrite shapes with different $S$ from axisymmetric simulations at a fixed isotherm velocity $V_{\mathrm{iso}}$ = 0.12 m/s. (c) Dendrite shapes at different time steps indicating the tip burgeoning instability in an axisymmetric simulation with $S=5$ at a velocity close to 1.64 m/s. All simulations were performed for the solidification of an Al-3wt.\% Cu alloy with $G = 5 \times 10^6$ K/m.
\label{fig:PF_axissym}
}
\end{figure}

3D dendrites arranged in an extended periodic array can be approximated by a single dendrite in a cylindrical tube with an axisymmetric shape. In perfectly periodic states, such as hexagonal and cubic arrays, planes of symmetry align with the grooves encircling a cell. Consequently, the steady-state solution for the entire array can be derived by solving the problem for a single cell within a prism, applying reflection boundary conditions on its sides. This solution is well approximated by the one inside a cylinder, as illustrated in Fig.~\ref{fig:PF_axissym}(a). Given that axisymmetric simulations are intrinsically 2D, they perform much faster than full 3D simulations. This efficiency enabled us to systematically explore varying interface thicknesses, ranging from $S=5$ to 1. In the axisymmetric approximation \cite{gurevich_phase-field_2010}, the free-energy functional can be written as the integral with respect to $r$ and $z$ in the cylindrical coordinates as shown in detail in Appendix \ref{Appendix:Axisymm_PF}.

We first examined the steady-state dendrite shapes with different $S$ values from axisymmetric simulations at an isotherm velocity of $V_{\mathrm{iso}}=0.12$ m/s, as shown in Fig.~\ref{fig:PF_axissym}(b). This comparison reveals that different $S$ values produce almost identical dendrite shapes, and the computation time is reduced by three orders of magnitude for $S=5$ compared to $S=1$. Then we utilize $S=5$ for computational efficiency to simulate the cellular-dendritic to banding transition. By slowly increasing $V_{\mathrm{iso}}$ over an extended timescale—far surpassing the characteristic time required for the interface to stabilize into a steady-state shape—we could probe pattern stability across a broad range of dendrite growth rate. As shown in Fig.~\ref{fig:PF_axissym}(c), an instability similar to the 2D simulations presented in \cite{ji_microstructural_2023} is observed, which is highly localized at the dendrite tip and triggers a rapid ``burgeoning-like'' growth of the interface. 

Nevertheless, it is important to note that the axisymmetric simulation is carried out for an axisymmetric shape that is independent of the azimuthal angle $\varphi$ in a plane parallel to the growth direction, and the anisotropy functions are approximated by averaging azimuthal directions (Appendix \ref{Appendix:Axisymm_PF}). This simulation does not investigate systematically the role of the dendrite's anisotropic shape that is a function of both $\varphi$ and the polar angle $\theta$. Furthermore, this approximation assumes a single dendrite in a cylindrical tube with an axisymmetry. Hence, we cannot apply this approach for the simulation of planar interface including the banded microstructure, where the axisymmetry is broken. For these reasons, we also carry out full 3D PF simulation.

\subsubsection{Full three-dimensional simulations}

\begin{figure}[htbp!]
\includegraphics[scale=1]{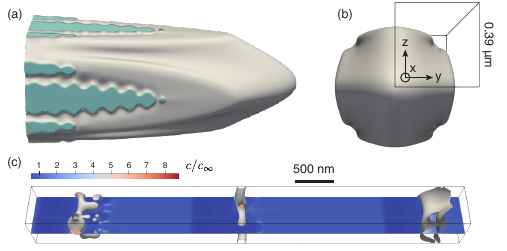}
\centering
\caption{
Fully 3D PF simulation for the solidification of an Al-3wt.\% Cu alloy with $V_{\mathrm{iso}}=0.48$ m/s and $G = 5 \times 10^6$ K/m: (a) side view and (b) top view. The colored regions represent the connected parts between dendrites. The black box in (b) indicates the simulation domain. (c) The banded microstructure at $V_{\mathrm{iso}}=0.24$ m/s, where the cross-section represents the $c/c_{\infty}$ colormap and the 3D contours represent the solid-liquid interface at $\phi=0$. \label{fig:3D_PF}
}
\end{figure}

The growth of the 3D dendritic array is influenced by the interplay of growth kinetics and anisotropic interface properties. Both of them have a four-fold symmetry in our consideration for the Al-Cu alloy here, with the anisotropy functions given in Eqs.\ \eqref{as_3D}-\eqref{ak_3D}. Our spatially extended 3D PF simulations of rapid solidification show that a cubic dendritic array structure is selected with these anisotropies. Thus, we utilize the reflection symmetry of a cubic array and model only a quarter of the dendrite in a $3.06 \times 0.39 \times 0.39$ $\mathrm{\mu m^3}$ simulation domain shown in Fig.~\ref{fig:3D_PF}, with reflection (no-flux) boundary conditions in the $\pm y$ and $\pm z$ directions. This simulation geometry corresponds to a primary spacing $\Lambda = 0.78$ $\mu$m, which is located within the stable range of $\Lambda$. The boundary conditions in the $\pm x$ directions are the same as the 2D simulations in Sec.~\ref{Sec:2D_dendrite_shapes}. The initial condition of the full 3D simulation of dendritic array is the SSD solution from the axisymmetric simulation with the same $\Lambda$ and under the same growth conditions. We interpolate the $\phi$ and $\tilde{c}$ fields within a cylindrical tube using the axisymmetric simulation results and extrapolate their values outside of this cylindrical tube in the 3D cuboidal simulation domain.

\begin{figure}[htbp!]
\includegraphics[scale=1]{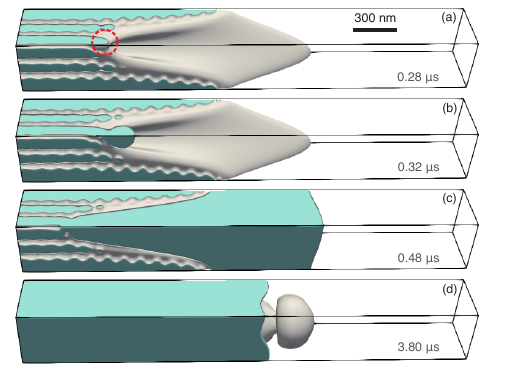}
\centering
\caption{
Evolution of the solid-liquid interface in a fully 3D PF simulation illustrating the instabilities at $V_{\mathrm{iso}}=V_{c,2}=0.816$ m/s for Al-3wt.\% Cu and $G = 5 \times 10^6$ K/m. A movie corresponding to this simulation is shown in the Supplemental Material \cite{SuppLongAM}. \label{fig:3D_PF_instability}
}
\end{figure}

As shown in Fig.~\ref{fig:3D_PF}(a)-(b), the steady-state 3D dendritic growth exhibits an elongated primary trunk with four arms branching out in the four energetically favored directions. As the growth rate increases, the dendrite becomes unstable. Here, the transition from dendritic to banding is caused by the tail instability originating from the connected part in grooves nestled between dendritic arms and normal to the growth direction, as shown in Fig.~\ref{fig:3D_PF_instability} and the Supplementary Movie \cite{SuppLongAM}. When the velocity approaches a critical value $V_{c,2} = 0.816$ m/s, the growth of the tail, shaped like a corrugated roof, abruptly accelerates and expands laterally, surpassing the dendritic tip. The location where the instability occurs is different from the observations in the axisymmetric simulations, where the burgeoning instabilities are highly localized at the dendrite tip region. In the confined axisymmetric simulations, the periodic deep liquid grooves separating the dendritic array are extremely narrow, typically preventing the emergence of tail instability. However, in the context of a full 3D dendrite with its four arms, the liquid grooves between these arms offer ample space, facilitating the development of tail instability. A quite similar situation is also observed in the spatially extended 2D simulations of banding, as shown in Supplementary Movie 2 of Ref.~\cite{ji_microstructural_2023}. During the slow growth of the dendritic segment corresponding to the ``dark'' band, an abrupt acceleration of the tail occurs beneath the solidification front. This tail rapidly develops and expands laterally to overtake dendritic tips, leading to the formation of the microsegregation-free ``light'' band.
As the planar solidification front of the light band decelerates and the interface undercooling increases, the interface becomes unstable again. As shown in Fig.~\ref{fig:3D_PF_instability}(d), a burgeoning instability triggers a transition from planar to dendritic growth during a banding cycle. These planar-to-dendritic cycles persist, forming banded microstructures, as shown in Fig.~\ref{fig:3D_PF}(c).

\begin{figure}[htbp!]
\includegraphics[scale=0.6]{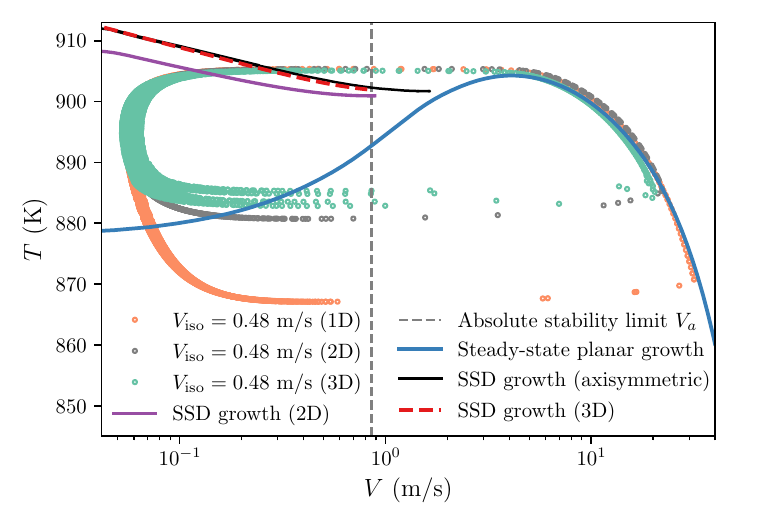}
\centering
\caption{
Steady-state curves and banding cycles in the T-V plane. Simulations are performed for Al-3wt.\% Cu alloy with $G = 5 \times 10^6$ K/m. \label{fig:T_V}
}
\end{figure}

To further investigate the stability of SSD solutions, we first performed axisymmetric and full 3D simulations at an $V_{\mathrm{iso}}$ below the onset of banding to form a steady-state dendrite array structures. It was then followed by slowly increasing $V_{\mathrm{iso}}$ in time on a timescale much longer than the characteristic time for the interface to relax to a steady-state shape. As illustrated in Fig.~\ref{fig:T_V}, the SSD in the full 3D simulation closely mirrors that of the axisymmetric simulation in the $T$-$V$ plot, up until a critical isotherm velocity $V_{c,2} \approx 0.82$ m/s where the tail instability is triggered. This $V_{c,2}$ closely approaches the absolute stability limit $V_a \approx 0.86$ m/s according to Eq.~\eqref{abs_stab} with $k(V)$ and $m(V)$ computed from the approximate PF solution in 1D, and assuming $\Gamma = \Gamma_0$ for simplicity. In 2D simulations, the critical velocity $V_{c,2} \approx 0.88$ m/s was found to be close to $V_a$ where a highly localized instability at the dendrite tip occurs \cite{ji_microstructural_2023}. The proximity of $V_{c,2}$ to $V_a$ in both 2D and 3D simulations might not be coincidental because the analysis for $V_a$ is for a planar interface, and the connected part in the tail of the 3D simulation is growing in the normal direction, similarly to the burgeoning instability in 2D where the tip develops into a corrugated roof. In comparison, the axisymmetric SSD survives beyond $V_a$ and extends toward the steady-state planar growth curve in the $T$-$V$ diagram (see Fig.~\ref{fig:T_V}), which is the characteristic solidus temperature $T(V)$ for a planar interface. The axisymmetric SSD terminates at a higher critical velocity $V_{c,2} \approx 1.64$ m/s, where a tip burgeoning instability occurs. The inherent constraints of axisymmetric conditions force the instabilities to assume the shape of hemispherical caps instead of the corrugated patterns seen in 2D and 3D simulations. This distinction underlies the significant discrepancy in $V_{c,2}$ observed for axisymmetric simulations.

Beyond the critical isotherm velocity $V_{c,2}$, we observe the formation of banded microstructures, as depicted in the initial banding cycle of Fig.~\ref{fig:3D_PF_instability}. Our simulations indicate the existence of bi-stable solutions—banding or SSD—influenced by the initial conditions within a specific velocity range. For $V_{\mathrm{iso}} \ge 0.24$ m/s, alternating banded patterns emerge, as demonstrated in Fig.~\ref{fig:3D_PF}(c), when initiating from a stationary planar interface at the liquidus temperature. An SSD solution is also achievable at these velocities by incrementally increasing $V_{\mathrm{iso}}$ from an existing SSD solution at lower velocities. The observed 3D band spacing is approximately 2~$\mu$m, consistent with 2D simulations using the FTA, yet substantially exceeds the experimentally observed spacing of about 500~nm. A difference between 2D and 3D simulations is the prominence of the light bands in 3D, in contrast to 2D simulations where dark bands can be more dominant, as shown in Fig.~4(a) of Ref.~\cite{ji_microstructural_2023}. This difference is likely due to the spatially extended dendritic growth within larger simulation domains, which promotes the formation of dark bands but also increases computational demands. To quantitatively capture the banded microstructure, integrating TFC in 3D and using a larger simulation domain size are necessary, which we reserve for future work. In Fig.~\ref{fig:T_V}, we also plot the $T$-$V$ oscillation cycles across 1D, 2D, and 3D simulations, all conducted under identical growth conditions at $V_{\mathrm{iso}}=0.48$ m/s. Despite the larger cycles in the 1D planar interface simulations, the 2D and 3D simulations yield similar cycles, with dendritic to planar transitions at higher temperatures around 885 K. During deceleration, where the interfaces maintain planar, all simulations adhere to the 1D steady-state planar growth trajectory (solid blue curve in Fig.~\ref{fig:T_V}).

\section{Conclusions and perspectives} \label{Sec:Conclusions_perspectives}
 

In this study, we provided a comprehensive derivation and analysis of quantitative PF models for rapid solidification. These models utilize upscaled interface thicknesses and reproduce quantitatively solute trapping effects across a broad interface velocity range, $V$, spanning six orders of magnitude from micron/s to m/s. This advance addresses the dual challenges of carrying out simulations on experimentally relevant length and time scales and incorporating nonequilibrium effects at the solid-liquid interface, enabling the quantitative simulations of microstructural pattern formation under far-from-equilibrium conditions. The generic variational framework presented herein also allows for extending the PF simulations to incorporate additional complexities, such as thermal diffusion and the complex free-energy functions for general binary alloys included in this work. Using the proposed PF models, we conducted an array of simulations in both 2D and 3D to explore dendrite shape convergence, microstructure selection, and the transition of cellular/dendritic to planar/banding.

Building on Model I from Ref.~\cite{ji_microstructural_2023}, we derived model equations variationally from a free-energy functional for a dilute binary alloy. A phenomenological free-energy is chosen such that the stationary solution of the PF profile $\phi_0(x)$ is exactly a hyperbolic tangent function \cite{karma_phase-field_2003}. Asymptotic analysis in the larger $V$ limit with the $\phi(x)=\phi_0(x)$ assumption demonstrated that, this PF model in its $S=1$ scenario matches exactly to the sharp-interface CG model with a solute drag coefficient $\alpha \approx 0.645$ for a given $g(\phi)$ function. A quadratic form of $q(\phi)$ has been adopted to enhance solute diffusivity for $S>1$, thereby eliminating excess solute trapping perpendicular to the interface. This allowed us to carry out quantitative simulations with $W \gg W_0$. Solving the PF equations in 1D demonstrated that the model reproduces quantitatively $k(V)$ and $m(V)$ curves over a broad range of $V$ for different $S$. Although the agreement between the PF and CG models are not generally expected for the entire $V$ range, we found their agreement is almost perfect for large $k_e$. Additionally, we examined the ramifications of excess surface diffusion (due to heightened solute diffusivity at the interface) and introduced a modified PF model that separates solute diffusion in normal and tangential directions. Though the excess surface diffusion exhibits minimal effects in 2D dendrite growth simulations, its combined effects with interface stretching at higher velocities require further investigation.

While Model I is effective in simulating a wide range of dilute binary alloy systems, its application to cases with higher nominal concentrations $c_{\infty}$, smaller $k_e$, or increased interface thicknesses may induce numerical issues at certain interface driving forces, leading to the emergence of unphysical phases. Similar issues are expected to arise in other PF models \cite{kim_phase-field_1999,plapp_unified_2011}. To address these issues, we introduced Model II that interpolates the free energy of two bulk phases at the interface, which is equivalent in its $S = 1$ case to PF models in Refs.~\cite{wheeler_phase-field_1992, warren_prediction_1995}. Model II reproduces the solute trapping and quantitative $k(V)$ and $m(V)$ curves similar to those of Model I for dilute alloys. While Model II is more robust—avoiding the aforementioned numerical issues for all materials and modeling parameters and providing simulation results similar to Model I—it requires stricter conditions for numerical stability. Thus, we identify safe operational regions in the $\tilde{\lambda}$-$k_e$ parameter space for Model I and exclusively use it therein. Model I also serves as the basis for building the variation models, including the model without excess surface diffusion and the axisymmetric model. 

Model II also facilitates the modeling of non-dilute binary alloys by integrating complete free-energy functions for each bulk phase, typically resulting in nonlinear liquidus and solidus. This integration is crucial for the accurate modeling of solidification processes in concentrated alloys. The 1D PF solutions, which integrate free-energy functions of Al-Ag alloys, show that the model produces liquidus and solidus curves at low $V$ closely match the equilibrium phase diagram derived from CALPHAD. Moreover, nonequilibrium solutions at high $V$ indicate that the liquidus and solidus shift toward the $T_0^E$ line, representing equal free energy between the two phases at equilibrium. This shift is a manifestation of the solute trapping effects that become significant at higher $V$. The agreement between the nonequilibrium phase diagrams predicted by the PF model for $S=1$ and $S=5$ confirms the effectiveness of the enhanced diffusivity approach with an upscaled interface for concentrated alloys. Therefore, this study provides a robust pathway for quantitatively modeling the rapid solidification of general binary alloys with complex phase properties. Additionally, the PF approach could also serve as a valuable tool for calculating the nonequilibrium phase diagram.

Microstructure selection maps were generated using 2D PF simulations under both FTA and TFC conditions. For very dilute binary alloys, a smooth transition from cellular/dendritic to planar front growth is observed at a critical isotherm velocity $V_c$. However, for solute concentrations above a critical value $c^*$, the transition from cellular/dendritic to banding occurs at critical isotherm velocities ranging from $V_{c,1}$ to $V_{c,2}$, depending on the initial conditions. Our 2D simulations highlighted the strong dependence of $c^*$ on thermal conditions, revealing a higher $c^*$ associated with latent-heat diffusion consistent with the predictions of linear stability analysis \cite{karma_dynamics_1992,karma_interface_1993}.
Furthermore, we demonstrate that the values of $V_c$ and $V_{c,2}$ predicted by the PF simulations with $S=1$ closely match the absolute stability limit across various alloy compositions. This agreement suggests that the analytical theory on the stability of dendritic solutions is well reproduced by the PF model. For PF simulations with $S>1$, the predictions of the critical isotherm velocities converge to those with $S=1$, except for the prediction of $V_{c,2}$ for lower concentrations near $c^*$. This convergence issue is likely due to excess surface diffusion and interface stretching for $S > 1$. We have verified that eliminating excess surface diffusion alone is insufficient to resolve this issue. A more comprehensive study of the individual and combined effects of excess surface diffusion and interface stretching is needed in future work.

In our investigation of 3D microstructural patterns, both axisymmetric approximations and full 3D simulations were performed. Axisymmetric simulations were used to approximate 3D dendrites, revealing that nearly identical dendrite shapes can be obtained in simulation using different interface thickness from $S=5$ to 1. The full 3D simulations revealed an instability at the tail of a 3D dendrite, which occurs at a critical velocity $V_{c,2}$ very close to the absolute stability limit $V_a$. The agreement between $V_{c,2}$ and $V_a$ is also observed in 2D, where this instability manifests as a corrugated roof \cite{ji_microstructural_2023}.
These results suggests that the standard theory of absolute stability reasonably predicts the upper critical velocities for the onset of instability in both 2D and fully 3D dendrites, despite they have different morphological manifestations of this instability.
Additionally, banded microstructures were observed in the full 3D simulations, with their oscillation cycle in the $T$-$V$ plane similar to that in 2D simulations.

Apart from this work and that in \cite{ji_microstructural_2023}, there have been other efforts to extend PF modeling to rapid solidification. Prior PF formulations have been limited to modeling a small departure from equilibrium \cite{pinomaa_quantitative_2019,pinomaa_phase_2020}, or have only reproduced solute trapping in 1D setups at higher $V$ \cite{ahmad_solute_1998,karma_phase-field_2003,danilov_phase-field_2006,galenko_solute_2011,steinbach_phase-field_2012,kavousi_quantitative_2021,mukherjee_quantitative_2023}. However, to quantitatively model the dendritic and banded microstructures in 2D and 3D, especially their transitions in the vicinity of the absolute stability limit, a PF model has to overcome the aforementioned dual challenges. Our work provides a general PF framework that not only addresses these challenges but also facilitates straightforward numerical implementation. Owing to the intrinsic capture of solute trapping effects within the proposed PF model, only a limited number of input parameters or functions are needed, predominantly those associated with physical properties of alloys. Hence, our approach offers practical and effective tools for quantitative PF modeling of rapid solidification, thereby enabling accurate predictions of microstructural pattern formation under conditions relevant to real-world solidification processes.

This work may be expanded in several directions. Firstly, one limitation of the present model is that, like the CG model, it does not capture the fact that complete solute trapping [$k(V)=1$] occurs at a finite velocity as demonstrated by atomistic simulations \cite{yang2011atomistic}. This effect is captured by alternate models of rapid solidification that relax the assumption that the diffusive flux relaxes instantaneously to a steady-state \cite{galenko1997local,galenko2007solute,galenko_solute_2011}. One interesting future prospect would be to extend the present model formulation to relax this assumption in order to more accurately model solute trapping in the velocity range where complete trapping occurs. It is worth noting, however, that at velocities where complete trapping is predicted to occur (e.g., Fig.~2 in \cite{yang2011atomistic}), which are much larger than the diffusive speed, the CG model and present PF model predict a slightly lower amount of trapping as $k(V)$ is already close to unity in this range. Therefore, we do not expect the microstructural predictions of our present model including the banding cycles to be strongly affected by the weak partitioning that persists for velocities much larger than the diffusive speed. This expectation is supported by the fact that the present model has yielded quantitative predictions of band spacing \cite{ji_microstructural_2023} and lateral spreading velocity of microsegregation-free bands \cite{ji2024microstructure}, both in good agreement with experiments. 
Secondly, the PF model can be adapted to model grain growth with multiple orientations, enabling the simulation of grain texture selection in rapidly solidified thin films. Additionally, quantitative 3D modeling that includes multi-physics can be applied to investigate 3D solidification processes, such as those relevant to metal additive manufacturing. It should be noted that the 3D simulations presented in this paper do not incorporate latent-heat diffusion, which is expected to significantly influence microstructural pattern formation beyond the absolute stability limit. For example, integrating TFC could enable the modeling of lateral growth during banding, as well as grain competitions due to different lateral spreading directions in 3D.
Furthermore, the integration of fluid dynamics could be applied for modeling the formation of microstructures within melt pools during the laser bed powder fusion process. The flexibility of the proposed variational PF framework also allows for its extension to multi-phase or multi-component alloys with integrating complete phase diagram information from CALPHAD. This positions it as a promising tool for quantitative modeling of rapid solidification and predicting solidification microstructures across various alloy systems.

\section*{Acknowledgement}

This work was supported by the U.S. Department of Energy (DOE), Office of Science, Basic Energy Sciences (BES) under Award No.~DE-SC0020895. K.J. acknowledges that this work was partially supported by Lawrence Livermore National Laboratory (LLNL) under Contract DE-AC52-07NA27344. We thank Longhai Lai for providing the Al-Ag phase diagram. Most numerical simulations were performed on the Discovery cluster of Northeastern University located in Massachusetts Green High Performance Computing Center (MGHPCC) in Holyoke, MA. Computing support for part of this work came from the LLNL Institutional Computing Grand Challenge program.

\appendix
\renewcommand{\thefigure}{A.\arabic{figure}}
\setcounter{figure}{0}  

\section{Variational analysis of the phase-field model without excess surface diffusion} \label{Appendix:Surf_diff}

We perform a variational analysis for the PF model in Sec.~\ref{Sec:surface_diffusion} that separates solute diffusion into components normal and tangential to the interface. Although it has been demonstrated that Eqs.\ \eqref{functional_dev_c}-\eqref{functional_dev_p} adhere to the gradient dynamics \cite{karma_phase-field_2003}, which guarantee the monotonic decrease of $F$ over time, a different variational analysis is necessary for the PF model with separated solute diffusion. Considering the same dynamics as in the main text, the equations are reiterated as
\begin{align}
\frac{\partial \phi}{\partial t} &= -K_{\phi} \frac{\delta F}{\delta \phi} \\
\frac{\partial c}{\partial t} &= \vec{\nabla} \cdot\left[K_{\bot}(\phi,c) \vec{\nabla}_{\bot} \mu^c+K_{\parallel}(\phi,c) \vec{\nabla}_{\parallel} \mu^c \right]
\end{align} 
Subsequently, the time derivative of the free energy functional yields
\begin{align}
\frac{d F}{d t}&=\int d \vec{x}\left(\frac{\delta F}{\delta \phi} \partial_t \phi+\frac{\delta F}{\delta c} \partial_t c\right) \\
&= -K_{\phi} \int d \vec{x}\left(\frac{\delta F}{\delta \phi}\right)^2 + I,
\end{align}
with
\begin{equation}
I \equiv \int d \vec{x} \mu^c \vec{\nabla} \cdot\left(K_{\parallel} \vec{\nabla}_{\parallel} \mu^c +K_{\perp} \vec{\nabla}_{\perp} \mu^c \right).
\end{equation}
Using the identity $A \vec{\nabla} \cdot \vec{B}=\vec{\nabla} \cdot(A \vec{B})-\vec{B} \cdot \vec{\nabla} A$ and the relation $(\vec{\nabla}_{\parallel} \mu^c) \cdot (\vec{\nabla}_{\bot} \mu^c) = 0$, we obtain
\begin{align}
I=&\int d s \mu^c \mathbf{n}_s \cdot\left(K_{\parallel} \vec{\nabla}_{\parallel} \mu^c+K_{\perp} \vec{\nabla}_{\perp} \mu^c \right) \nonumber \\
&- \int d \vec{x} \vec{\nabla} \cdot\left(K_{\parallel} \left| \vec{\nabla}_{\parallel} \mu^c \right|^2+K_{\perp} \left | \vec{\nabla}_{\perp} \mu^c \right|^2 \right),
\end{align}
where the surface integral in the first term on the right-hand-side is derived from a volume integral using the divergence theorem, and $\mathbf{n}_s$ denotes the outward pointing unit vector normal to a closed surface. Assuming there is no solute flux through the boundaries of the system, the surface integral equals to zero. Given that $K{\phi}$, $K_{\parallel}$, and $K_{\bot}$ are all non-negative, we conclude
\begin{equation}
\frac{d F}{d t} \leq 0.
\end{equation}
Therefore, the PF model with separation of solute diffusion also adheres to the gradient dynamics.

\section{Numerical implementation of thermal field calculation} \label{Appendix:TFC}

\begin{figure}[htbp!]
\includegraphics[scale=0.6]{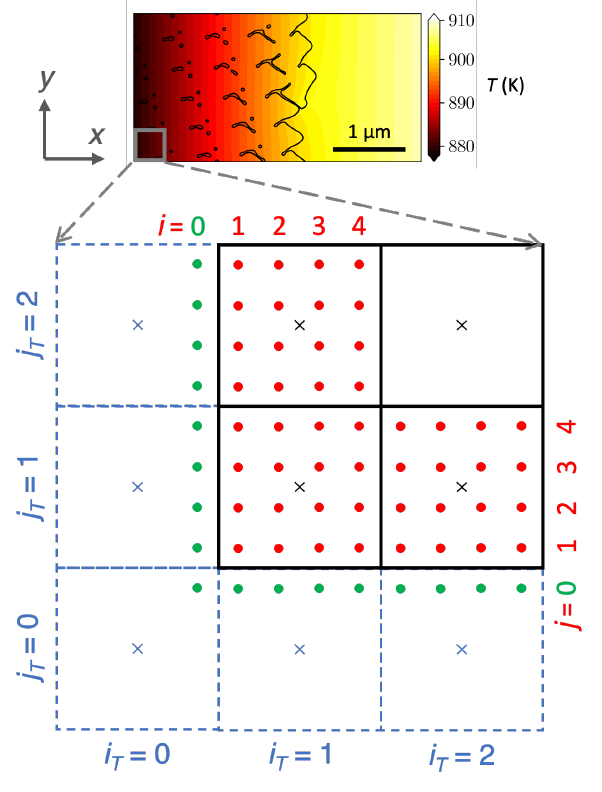}
\centering
\caption{
The finite-different grids in a 2D PF simulation with TFC. The PF and concentration field are solved on a finer grid indicated by dots; the thermal field is solved on a coarser grid indicated by crosses. The grid indexes $i,\,j=0$ and $i_T,\,j_T=0$ correspond to the boundary layers of the finer and coarser grids, respectively. The example simulation is performed for an Al-3wt.\% Cu alloy with $V_{\mathrm{iso}}=0.96$ m/s. \label{fig:TFC_grid}
}
\end{figure}

The evolution equations \eqref{Diemensionless_phi}-\eqref{Diemensionless_c} for $\phi$ and $\tilde{c}$ are solved on a 2D rectangular lattice using the finite difference method for spatial derivatives and an Euler explicit time-stepping scheme. The grid spacing and time step are denoted as $\Delta x$ and $\Delta t$, respectively. Due to the faster thermal diffusion compared to solute diffusion, the thermal equation \eqref{heat_diff} demands a more stringent numerical stability condition. In 2D, this condition is given by
\begin{equation}
\Delta t_T < \frac{\left(\Delta x_{T}\right)^{2}}{4 \widetilde{D}_{T}}, \label{dt_T}
\end{equation}
where $\Delta t_T$ and $\Delta x_{T}$ represent the time step and grid spacing for the thermal field, scaled by $\tau_0$ and $W$, respectively. With $\widetilde{D}_{T}/\widetilde{D}_{l} \sim 10^4$ for the alloy systems examined in this paper, using the same lattice with a grid spacing of $\Delta x$ reduces the time step by four orders of magnitude, rendering the numerical simulation nearly infeasible. A solution developed in Ref.~\cite{song_thermal-field_2018} employs a coarser grid $\Delta x_{T}>\Delta x$ for the thermal field, making the time step constant as $\Delta t = \Delta t_T$. This, however, results in the spatial resolution for the thermal field being approximately 100 times coarser than for $\phi$ and $\tilde{c}$. In this work, we introduce an approach that employs both a coarser grid and a finer time step to solve the coupled evolution equations, with both $\Delta x_{T}/\Delta x$ and $\Delta t/\Delta t_T$ being integers greater than one. For instance, in the Al-3wt.\% Cu simulations of Fig.~\ref{fig:selection_map}(b), we set $\Delta x_{T}/\Delta x=10$ and find the smallest ratio $\Delta t/\Delta t_T$ that meets the constraint in Eq.~\eqref{dt_T}, which is 3. Owing to the faster computation of the thermal equation (due to fewer grid points), this approach ensures efficient calculations for the coupled evolution equations, while retaining reasonable spatial resolution for the thermal field.

With the coarser-grid and multi-step strategy, we discretize the thermal equation \eqref{heat_diff} in 2D as:
\begin{align}
& \widetilde{T}_{t+\Delta t_T} = \widetilde{T}_{t} + \Delta t_T \left( \widetilde{V}_{\mathrm{iso}} \partial_{x} \widetilde{T} + \widetilde{D}_T \nabla^2 \widetilde{T} \right) \label{TFC_2D_dis} \\
& + H \sum_{x}^{x+\Delta x_{T}} \sum_{y}^{y+\Delta x_{T}}
\left[ \frac{\phi_{t+\Delta t}^{(x, y)}-\phi_{t}^{(x, y)}}{2} \right] \left( \frac{\Delta t_T}{\Delta t} \right) \left( \frac{\Delta x}{\Delta x_{T}} \right)^2 \nonumber
\end{align}
The double summation runs over all finer grid points $(i,j)$ within a region defined by a single coarser grid point $(i_T,j_T)$, as depicted in Fig.~\ref{fig:TFC_grid}. The first spatial derivative in Eq.~\eqref{TFC_2D_dis} is discretized as $\partial_{x} \widetilde{T}=[\widetilde{T}^{(x+\Delta x_T)}-\widetilde{T}^{(x-\Delta x_T)}]/(2 \Delta x_T)$, and the Laplacian is approximated using an isotropic form \cite{ji_isotropic_2022}. Between successive time steps for PF and concentration equations, Eq.~\eqref{TFC_2D_dis} is solved for $\Delta t/\Delta t_T$ steps. A similar discretization can be extended to a 3D cubic lattice using this coarser-grid and multi-step strategy.

During each main iteration with a step $\Delta t$, $\widetilde{T}$ is obtained at each point $(i,j)$ on the finer grid through bilinear interpolation of the $\widetilde{T}$ field, using the four nearest points on the coarser grid. Subsequently, this interpolated $\widetilde{T}$ is substituted into Eq.~\eqref{Diemensionless_phi} for $\phi$ and $\tilde{c}$ computation. While Eqs.\ \eqref{Diemensionless_phi}-\eqref{Diemensionless_c} are derived in the material frame, Eq.~\eqref{heat_diff} is solved in an adiabatic zone within the moving frame, which requires the location tracking (in the $x$ direction) of the simulation domain in the moving frame.
This location information of the finer grid is used for both numerical integration and interpolation. Typically, the adiabatic zone's size is much larger than the simulation domain for $\phi$ and $\tilde{c}$. For all TFC simulations in this study, we consider an adiabatic zone with a length 50 $\mu$m in the $x$ direction, with temperatures at the $\pm x$ boundaries set at fixed values 1001.4 K and 750.0 K, respectively. This thermal condition corresponds to a typical melt pool in a thin film experiment with a radius of 50 $\mu$m and a temperature gradient of 5 K/$\mu$m. The $\pm y$ boundary conditions of the thermal field follow those of $\phi$ and $\tilde{c}$, which are periodic unless specified otherwise.

\section{Full phase-field solution} \label{Appendix:Full_PF_solution}

The evolution equations \eqref{Diemensionless_phi}-\eqref{Diemensionless_c} are formulated within the material frame, wherein isotherms move at a constant speed $V_{\mathrm{iso}}$ in the $+x$ direction. To find stationary solutions of the PF model in 1D, these equations are transformed into the moving frame, where the isotherms are stationary. This leads to the following transformations
\begin{align}
\frac{\partial \phi}{\partial t} & \to \frac{\partial \phi}{\partial t} + \widetilde{V}_{\mathrm{iso}} \frac{\partial \phi}{\partial x}, \\
\frac{\partial \tilde{c}}{\partial t} & \to \frac{\partial \tilde{c}}{\partial t} + \widetilde{V}_{\mathrm{iso}} \frac{\partial \tilde{c}}{\partial x}, \label{lab_frame_c}
\end{align}
where $\tilde{c} \equiv c/c_{\infty}$ and $\widetilde{V}_{\mathrm{iso}} \equiv \tau_0 V_{\mathrm{iso}} /W $. Nevertheless, a direct computation of ${\partial_{x} \tilde{c}}$ in Eq.~\eqref{lab_frame_c} may induce numerical instability. Hence, we replace ${\partial_{x} \tilde{c}}$ using Eq.~\eqref{c_scale}, which implicitly assumes a stationary solution for the concentration profile. This ``enforced'' dynamics ensures that the 1D PF solution relaxes into a stationary state rather than oscillating. The dimensionless evolution equations for directional solidification in the moving frame are consequently given by
\begin{align}
\frac{\partial \phi}{\partial t}=& \widetilde{V}_{\mathrm{iso}} \frac{\partial \phi}{\partial x} 
 + (1+\epsilon_k) \partial_{xx} \phi + \frac{(1+\epsilon_k)}{(1+\epsilon_s)^2} \biggl \{ \phi-\phi^{3} \biggr. \nonumber \\
&\biggl. - \tilde{\lambda} g^{\prime}(\phi) \left[\tilde{c}+ \left( \frac{x-x_0}{\widetilde{l}_T}-1 \right) e^{b(1+g(\phi))}\right] \biggr \}, \label{eq_p_DS} \\
\frac{\partial \tilde{c}}{\partial t}=&\widetilde{D}_l \partial_x \left \{ q(\phi) \tilde{c} \partial_x [\ln \tilde{c}-b g(\phi)] \right \} \nonumber \\
&+ \widetilde{V}_{\mathrm{iso}} \left [ S (1-\tilde{c}) \frac{v}{q(\phi)} + b \tilde{c} \frac{d g(\phi)}{d x} \right ], \label{eq_c_DS}
\end{align}
where $v \equiv V_{\mathrm{iso}}/V_d^0$, and $x_0$ denotes the reference location in the moving frame, coinciding with the equilibrium liquidus temperature $T_L$. Here, we have considered the preferred growth direction in $x$, and both interface free-energy and kinetic anisotropies with the substitutions $S W_0 \rightarrow S W_0\left(1+\epsilon_s\right)$ and $\tau_0 \rightarrow \tau_0\left(1+\epsilon_k\right)$. In Eq.~\eqref{eq_c_DS}, the term $q(\phi)$ in the denominator approaches zero in the solid phase. To prevent the simulation from diverging, we regularize the term by adding a small quantity $10^{-8}$ in the denominator during the numerical simulation. By dynamically solving Eqs.\ \eqref{eq_p_DS}-\eqref{eq_c_DS} in the moving frame, the 1D $\phi$ and $\tilde{c}$ profiles relax to the stationary solutions for Model I. These dynamically derived 1D solutions can also be obtained by using a Newton solver. Additionally, we follow the same procedure to obtain the 1D $\phi$ and $\tilde{c}$ stationary solutions for Model II with dilute alloys.

The same strategy of replacing ${\partial_{x} \tilde{c}}$ with its stationary form and using the ``enforced'' dynamics to avoid the oscillatory solution can still be applied to Model II for concentrated alloys to obtain the full dynamical solutions with the integration of CALPHAD free-energy functions, which will be reported in a separate study.

\section{Axisymmetric phase-field model} \label{Appendix:Axisymm_PF}

In the axisymmetric approximation, the free-energy functional is represented as an integral over $r$ and $z$ in cylindrical coordinates:
\begin{equation}
F[\phi, c, T] = \int 2 \pi r F_v (\phi, \vec{\nabla} \phi, c) d r d z,
\end{equation}
where $F_v$ is the integrand from Eq.~\eqref{F_functional}. The evolution equation for the PF is then given by
\begin{align}
&\tau(\theta) \partial_t \phi(r, z, t)= -\frac{1}{2 \pi r} \frac{\delta F}{\delta \phi(r, z, t)} \nonumber \\
= & -\frac{\partial F_v}{\partial \phi}+\partial_z \frac{\partial F_v}{\partial\left(\partial_z \phi\right)}+\partial_r \frac{\partial F_v}{\partial\left(\partial_r \phi\right)}+\frac{1}{r} \frac{\partial F_v}{\partial\left(\partial_r \phi\right)},
\end{align}
where the last term can be expanded as
\begin{equation}
\frac{1}{r} \frac{\partial F_v}{\partial\left(\partial_r \phi\right)} = \frac{1}{r} \left(|\vec{\nabla} \phi|^2 W(\mathbf{n}) \frac{\partial W(\mathbf{n})}{\partial\left(\partial_r\phi\right)}\right)+\frac{1}{r} W^2(\mathbf{n})\partial_r\phi.
\end{equation}
Using the axisymmetric approximation \cite{gurevich_phase-field_2010}, the orientation-dependent interface thickness is described by
\begin{equation}
W(\mathbf{n}) = S W_0 a_s (\mathbf{n}) \approx S W_0 \bar{a}_s (\theta),
\end{equation}
where the anisotropy function $a_s (\mathbf{n})=a_s(\theta, \varphi)$ is approximated by averaging all azimuthal directions $\varphi$ as
\begin{align}
\bar{a}_s (\theta) & = \int_0^{2 \pi} \frac{d \varphi}{2 \pi} a_s (\theta, \varphi) \nonumber \\
& = (1-3 \epsilon_s)\left[1+\frac{4 \epsilon_s}{1-3 \epsilon_s}\left(\cos ^4 \theta+\frac{3}{4} \sin ^4 \theta\right)\right]. \label{as_axis}
\end{align}
Here, $\theta$ is the angle between the unit vector perpendicular to the interface pointing into the liquid and the growth axis $z$, given by
\begin{equation}
\theta=\arctan \frac{\partial_r \phi}{\partial_z \phi}.
\end{equation}
Similarly, the approximated kinetic anisotropy function $\bar{a}_k (\theta)$ takes the same form as in Eq.~\eqref{as_axis}.

Consequently, the evolution equation for the PF is
\begin{align}
\tau(\theta) \frac{\partial \phi}{\partial t}=& \vec{\nabla} \cdot\left[W(\theta)^{2} \vec{\nabla} \phi\right] +\phi-\phi^{3} \nonumber \\
&+ \sum_{i=r,z}\left[\partial_{i}\left(|\vec{\nabla} \phi|^{2} W(\theta) \frac{\partial W(\theta)}{\partial\left(\partial_{i} \phi\right)}\right)\right] \nonumber \\
&+ \frac{1}{r} \left(|\vec{\nabla} \phi|^2 W(\theta) \frac{\partial W(\theta)}{\partial\left(\partial_r \phi\right)}\right) + \frac{1}{r} W^2(\theta)\partial_r\phi \nonumber \\
&- \lambda g^{\prime}(\phi) \left[c+ \frac{\left(T-T_{M}\right)}{m_{e}} e^{b(1+g(\phi))}\right] \label{axis_eq_phi}
\end{align}
where $\tau(\theta)=\tau_0 \bar{a}_s(\theta)^2 / \bar{a}_k(\theta)$. In addition, the evolution equation of the concentration field becomes
\begin{align}
\frac{\partial c}{\partial t}=&\vec{\nabla} \cdot\left\{D_l q(\phi) c \vec{\nabla}[\ln c-b g(\phi)]\right\} \nonumber \\
&+\frac{1}{r} D_l q(\phi) c \partial_r [\ln c-b g(\phi)] \label{axis_eq_c}
\end{align}
Eqs.\ \eqref{axis_eq_phi}-\eqref{axis_eq_c} are solved numerically on a 2D square lattice similar to 2D simulations.

\paragraph*{Solving the anisotropy numerically}

The first anisotropy term in Eq.~\eqref{axis_eq_phi} can be expressed as
\begin{equation}
\vec{\nabla} \cdot\left[\bar{a}_s(\theta)^2 \vec{\nabla} \phi\right] = \bar{a}_s^2 \vec{\nabla}^2 \phi + \left(\partial_z \phi \partial_z \theta+\partial_r \phi \partial_r \theta\right) 2 \bar{a}_s^{\prime} \bar{a}_s,
\end{equation}
where we have scaled the length by $S W_0$. The second term expands as
\begin{align}
\sum_{m=z, r}&\left[\partial_m\left(|\vec{\nabla} \phi|^2 \bar{a}_s(\theta) \frac{\partial \bar{a}_s(\theta)}{\partial\left(\partial_m\phi\right)}\right)\right] \nonumber \\
&=\left(\partial_z \phi \partial_r \theta-\partial_r \phi \partial_z \theta\right) (\bar{a}_s^{\prime \prime} \bar{a}_s+\bar{a}_s^{\prime 2}),
\end{align}
and the third anisotropy term as
\begin{equation}
|\vec{\nabla} \phi|^2 \bar{a}_s(\theta) \frac{\partial \bar{a}_s(\theta)}{\partial\left(\partial_r \phi\right)} = \partial_z \phi \bar{a}_s^{\prime} \bar{a}_s
\end{equation}
where 
\begin{align}
\partial_z \theta = \frac{\partial_{z r} \phi \partial_z \phi-\partial_r \phi \partial_{z z} \phi}{|\vec{\nabla} \phi|^2}, 
\end{align}
and
\begin{equation}
\partial_r \theta = \frac{\partial_{r r} \phi \partial_z \phi-\partial_r \phi \partial_{z r} \phi}{|\vec{\nabla} \phi|^2}.    
\end{equation}

Therefore, the anisotropy function in Eq.~\eqref{as_axis} can be written as
\begin{align}
\bar{a}_s (\theta) = 1-3 \epsilon_s + \epsilon_s \frac{3 (\partial_r \phi)^4+4 (\partial_z \phi)^4}{|\vec{\nabla} \phi|^4}.
\end{align}
The first derivative of the anisotropy function is
\begin{align}
\bar{a}_s^{\prime} (\theta)&=-\frac{1}{2} \epsilon_s \left[2 \sin (2 \theta )+7 \sin (4 \theta )\right] \\
&=4 \epsilon_s \frac{[ 3 (\partial_z \phi)(\partial_r \phi)^3 - 4 (\partial_r \phi) (\partial_z \phi)^3]}{|\vec{\nabla} \phi|^4},
\end{align}
and its second derivative is
\begin{align}
\bar{a}_s^{\prime\prime} (\theta) &= -2 \epsilon_s \left[ \cos (2 \theta )+7 \cos (4 \theta )\right] \\
&=-4 \epsilon_s \frac{\left[ 3 (\partial_r \phi)^4 - 21 (\partial_r \phi)^2 (\partial_z \phi)^2 + 4 (\partial_z \phi)^4 \right]}{|\vec{\nabla} \phi|^4}.
\end{align}
Thus, $\bar{a}_s$, $\bar{a}_s'$, and $\bar{a}_s''$ are computed in their explicit forms in terms of the derivatives of $\phi$ during the axisymmetric simulation.

\paragraph*{Boundary conditions}

In the axisymmetric simulation, the boundary condition at $r=0$ needs to be treated carefully. We extrapolate $\phi$ at $r=0$ using the $\phi$ values at $r=\Delta x$ and $r=2\Delta x$. Using the following relations
\begin{align}
\phi_{r0} &\equiv \phi(r=0) = a_r, \\
\phi_{r1} &\equiv \phi(r=\Delta x) = a_r + b_r \Delta x^2, \\
\phi_{r2} &\equiv \phi(r=2\Delta x) = a_r + 4 b_r \Delta x^2,
\end{align}
where $a_r$ and $b_r$ are arbitrary constants, we obtain
\begin{equation}
\phi_{r0} = \frac{(4 \phi_{r1} - \phi_{r2})}{3}.
\end{equation}
Similarly, we obtain the boundary condition at $r=0$ for the concentration field as
\begin{equation}
c_{r0} = \frac{(4 c_{r1} - c_{r2})}{3}.
\end{equation}
The boundary condition in the $+r$ direction is no-flux, and the boundary conditions at $\pm z$ are consistent with the 2D simulations described in Sec.~\ref{Sec:2D_dendrite_shapes}.


\bibliography{references}

\end{document}